\documentclass[aps,prd,citeautoscript,nofootinbib, floatfix, notitlepage,twocolumn,nobibnotes]{revtex4-1}

\usepackage{amsmath,amsfonts,amssymb}
\usepackage{mathrsfs}
\usepackage{graphicx}
\usepackage[english]{babel} 
\usepackage{hyperref} 
\usepackage{verbatim}
\usepackage[usenames, dvipsnames]{color}
\usepackage{bbold}
\usepackage{overpic}

\newcommand{\QQ}{\mathbb Q}

\begin{document}

\title{Critical properties of the many-particle (interacting) Aubry-Andr\'e model ground-state localization-delocalization transition}

\author{Taylor Cookmeyer}
\email[]{tcookmeyer@berkeley.edu}
\author{Johannes Motruk}
\author{Joel E. Moore}
\affiliation{Department of Physics, University of California, Berkeley, CA, 94720, USA}
\affiliation{Materials Sciences Division, Lawrence Berkeley National Laboratory, Berkeley, California, 94720, USA}

\begin{abstract}
As opposed to random disorder, which localizes single-particle wave-functions in 1D at arbitrarily small disorder strengths, there is a localization-delocalization transition for quasi-periodic disorder in the 1D Aubry-Andr\'e model at a finite disorder strength. On the single-particle level, many properties of the ground-state critical behavior have been revealed by applying a real-space renormalization-group scheme; the critical properties are determined solely by the continued fraction expansion of the incommensurate frequency of the disorder. Here, we investigate the  many-particle
localization-delocalization transition in the Aubry-Andr\'e model with and without interactions. In contrast to the single-particle case, we find that the critical exponents depend on a Diophantine equation relating the incommensurate frequency of the disorder and the filling fraction which generalizes the dependence, in the single-particle spectrum, on the continued fraction expansion of the incommensurate frequency.  This equation can be viewed as a generalization of the resonance condition in the commensurate case.  When interactions are included, numerical evidence suggests that interactions may be irrelevant at at least some of these critical points, meaning that the critical exponent relations obtained from the Diophantine equation may actually survive in the interacting case.
\end{abstract}

\date{\today}

\maketitle
 
 \section{Introduction}
 
 The localization of a system around random disorder is a problem originally addressed by Anderson \cite{Anderson1958}. More recently, once interactions were added, such systems were shown to exhibit many-body localization (MBL) \cite{BaskoAleinerAltshuler2006,Gornyi2005,Abanin2019} whereby local integrals of motion prevent thermalization. Random disorder makes such systems difficult to study theoretically (due to the necessity of disorder-averaging) and experimentally (due to the challenge of engineering random disorder) \cite{vanNieuwenburg2019}. 
 
 Between random disorder and no disorder, there is quasi-periodic disorder as demonstrated by the Aubry-Andr\'e-(Harper) (AA) model \cite{Aubry1980,Harper1955}. 
 \begin{equation}
    H_{\rm AA} = \sum_i h_i \hat n_i -J(\hat c_i^\dagger \hat c_{i+1} + \hat c_{i+1}^\dagger c_i) \label{eq:H_AA}
\end{equation}
 with $h_i = \lambda J \cos(2\pi i \beta + \phi)$ for $\beta$ an irrational number.
 The model can also be understood as the result of a tight-binding square lattice Hamiltonian in the presence of a magnetic field yielding the famous Hofstadter butterfly when the hopping amplitudes are the same \cite{Hofstadter1976}. Within the single-particle spectrum, this model exhibits a 1D localization-delocalization transition at $\lambda=2$, which can be seen by considering the duality transformation $c_k = \sum_n \exp(2\pi i\beta kn) c_n / \sqrt{N}$ sending $\lambda\to4/\lambda$\cite{Aubry1980}.\footnote{Note that when $\beta$ is a Liouville number, the transition does not occur \cite{Kohmoto1983, Avron1982}, which should therefore be excluded when we say ``all $\beta$''.} 
 
 Adding the simplest interaction term $H_{\rm iAA} = H_{\rm AA}+\sum_i V n_i n_{i+1}$ leads to the interacting Aubry-Andr\'e (iAA) model.  The localization of the ground state was theoretically predicted to persist once interactions were included \cite{Mastropietro2015,Mastropietro2017}. Moreover, it was numerically demonstrated that the interacting model would exhibit many-body localization \cite{Iyer:2013Many-body, Khemani2017,Naldesi2016,Setiawan2017,Bera2017,Michal2014}, and the universal properties of the MBL transition were predicted to be different between the random and quasi-periodic case \cite{Khemani2017}, which has been shown in a toy model of MBL \cite{agrawal2019}. Interestingly, the MBL transition in the presence of interactions does not seem to exist close to $\lambda=2$ at $V\ll J$\cite{Znidaric2018}, but dynamical studies suggest that the MBL transition occurs at a large enough value of $\lambda$ and $V\sim J$ \cite{Doggen2019,Lev2017}. 

One of the great advantages of this model, as opposed to random disorder, is that it can be more easily realized experimentally both with interactions in cold atom systems \cite{Schreiber2015,Bordia2016,Luschen2017,Kohlert2019} and without interactions in cold atom systems \cite{Roati2008} and photonic lattices \cite{Lahini2009}. Experiments on the interacting model have so-far mostly focused on realizing the MBL transition \cite{Schreiber2015} and other aspects of MBL physics \cite{Bordia2016,Luschen2017,Kohlert2019}.
 
In addition to the fascinating higher-temperature properties of this system, the ground-state properties of this model, even in the free case, are still being explored. Recent work on the non-interacting model has focused mainly on dynamical studies \cite{Purkayastha2018,Varma2017,Saha2016,Wu2019,Sutradhar2019} or the critical properties of the transition \cite{Szabo2018,Wei2019}, and numerous generalizations of the model have been introduced to generate a system with a mobility edge \cite{Ganeshan2015,Li2015} or more topological features \cite{DeGottardi2013,Cai2013}. 

In this paper, we will focus on the ground state critical properties of the (interacting) Aubry-Andr\'e model where it was determined that $\nu=1$ for the $\lambda=2$ transition for all irrational $\beta$ \cite{Aubry1980}. It has been known that the universality class of the single-particle ground state depended solely on the continued-fraction expansion of $\beta$ \cite{Hashimoto1992}; only recently, however, with an explicit real-space renormalization group (RSRG) scheme \cite{suslov1982localization,Thouless1983,Azbel1979}, the authors of Ref.~\onlinecite{Szabo2018} derived an expression for the dynamic critical exponent, $z(\beta)$, for $\beta\ll 1$. Furthermore, as studied in Ref.~\onlinecite{Thakurathi2012}, a similar transition occurs in the limit $\beta = 1/q$ with $q \to \infty$ at half-filling. (The authors of Ref.~\onlinecite{Thakurathi2012} claim that $\nu\approx 0.7$ instead of the usual $\nu=1$, but we find that $\nu\approx 1$ later.)

The above RSRG scheme only works for the single-particle spectrum, and, although an RSRG for the middle of the spectrum exists \cite{Ostlund1984}, its assumptions are less physically clear, and the RSRG procedure depends strongly on $\beta$ and only works at certain fillings. There exists therefore an open question about $z$'s dependence on $\beta$ and $\rho$ where $\rho=N_F/N$ is the filling fraction. The value of $z$ at different filling fractions determines how different parts of the energy spectrum scale with system size \cite{Evangelou2000,Hiramoto1989,Cestari2011} and is therefore useful for understanding the multifractal properties of this system \cite{Tang1986} as well as the multi-particle ground-state transition properties. Furthermore, it determines the low-temperature specific heat \cite{Tang1986}. It was previously known that $z$ does depend on the filling fraction \cite{Hashimoto1992,Ostlund1984},  and it was incorrectly claimed that universality at half-filling was solely determined by the continued fraction expansion in Ref.~\onlinecite{Hashimoto1992} (likely because the study was limited to only certain $\beta$), 
but we are aware of no classification of the universality relation between different filling fractions and different $\beta$. 

  In this work, we will present such a classification scheme. The rational approximations originating from $\beta$'s continued fraction expansion defines a sequence of integers $M_k$ and $N_k$ such that $M_k/N_k\approx \beta$ (see below). We will present numerical evidence that the universality class is determined by the sequence of integer solutions $(Q_k,P_k)$ with $|Q_k|\le N_k/2$ as small as possible of the Diophantine equation  \begin{equation}\label{Diophantine}
     Q_k M_k - P_k N_k = \pm N_F
 \end{equation}
 where $N_F$ is the number of Fermions. Explicity, we conjecture that, at incommensurate fillings, the transition is at $\lambda_c=2$, $\nu=1$,  and $|Q_k|/N_k$ approaches a repeating sequence of $p$ values, which uniquely identify the universality class and thus a value for $z$. Therefore, as we will present, even at different fillings and different $\beta$, the universal properties of the transition can be the same. We additionally conjecture that, when $Q_k=q$ is fixed, at commensurate fillings, then the transition occurs at $\lambda_c=0$ with exponents $\nu=q$ and $z=1$.
 
 In the latter case, the Diophantine condition is simply the resonance condition between the perturbation at $k=2\pi\beta$ and $2k_F=2\pi \rho$, which occurs at $q$th order in perturbation theory for small $\lambda$. The Diophantine equation in the incommensurate case can be viewed as a generalization of that concept. Additionally, this same Diophantine equation has been considered for this model in other contexts such as the integer-quantum Hall effect as it is related to the Chern number of the band \cite{TKNN, ni2019,Kraus2012TopologicalEquivalence,Kraus2012TopologicalStates}. This understanding provides a framework for why there are different transitions within this model depending on the filling and the continued-fraction expansion of $\beta$. 
 
 Before we continue, let us consider the concrete example of the silver ratio, $\beta_a=\sqrt{2}-1$, and $\beta_b=1/(1+\beta_a)=1/\sqrt{2}$, in order to make the above statements more clear.  Both $\beta$'s continued fraction expansions are the same after the first term, and, in the single-particle spectrum, they would therefore be in the same universality class \cite{Hashimoto1992,suslov1982localization,Thouless1983,Azbel1979,Szabo2018}, which is predicted by our Diophantine equation conjecture (see Sec.~\ref{sec:discussion}). However, let us solve the Diophantine equation in the case of half-filling (i.e. $N_F=\lfloor N_k/2 \rfloor$) where a different result will emerge. 
 
 Recalling that the Pell numbers are $P_k=1,2,5,12,29,70,...$ where $P_{k+1}=2P_k+P_{k-1}$, the best rational approximations to $\beta_a$ are given by $M_k=P_{k}$ and $N_k =P_{k+1}$, while the best rational approximations to $\beta_b$ are given by $M_k=P_{k}$ and $N_k=P_k + P_{k-1}$. After specifying $M_k,N_k,$ and $N_F=\lfloor N_k/2\rfloor$, there are an infinite number of integer solutions $(Q_k,P_k)$ to Eq.~\eqref{Diophantine}, but we find the solution with $|Q_k|\le N_k/2$ for both $\pm N_F$, and, of those, we pick the solution with the smaller value of $|Q_k|$. We find that $|Q_k| =1,6,6,35,35,...=P_{2\lceil (k+1)/2\rceil}/2 $ for $\beta_a$ and $|Q_k| = 0,1,2,5,12,...=M_{k-1}$ for $\beta_b$. For $\beta_a$ and large enough $k$,  $|Q_k|/N_k=...,q_1,q_2,q_1,q_2,...$ where $q_1=1/2$ and $q_2=(\sqrt{2}-1)/2\approx M_k/(2N_k)$, whereas, for $\beta_b$ and large enough $k$, $|Q_k|/N_k=...,q_3,q_3,q_3,...$ with $q_3=1-1/\sqrt{2}$. 
 
 We would then conjecture that $\beta_a$ and $\beta_b$ belong to two different universality classes at half-filling, which is demonstrated in Fig.~\ref{fig:SR} below.  The periodicity of the values of $|Q_k|/N_k$ corresponds to the two universal curves for $\beta_a$ and the one universal curve for $\beta_b$.

 If we consider instead the golden ratio, $(\sqrt{5}-1)/2$ and  other $\beta$'s with the same asymptotic continued fraction expansion, they will ultimately have the same repeating part of the sequence of $|Q_k|/N_k$ with a periodicity of three at half-filling. Our Diophantine equation conjecture would then predict that they are in the same universality class, as was seen by \cite{Hashimoto1992} and as we observe numerically (see Table~\ref{tab:exponents} and Fig.~\ref{fig:GR}).

 Once interactions are turned on, the ground-state phase diagram becomes richer \cite{Naldesi2016} (see Ref.~\onlinecite{Roux2008} for the bosonic version), but the critical exponent $\nu$ of the localization-delocalization transition does not seem sensitive to the interaction strength at half-filling \cite{Schuster2002}. Having the same value for all $\nu=1$, it is an open question whether the dynamic critical exponent $z$ remains the same, which would suggest that the universality class is insensitive to the interaction strength. In the integer quantum Hall effect, the value of $\nu$ appears to be the same as the non-interacting model, but the value of $z=1$ seen in experiment is different than the $z=2$ predicted by the non-interacting model (see for instance Ref.~\onlinecite{lee1996} and references therein). 
 We find that the interaction does not change the exponent $z$, which suggests that the Diophantine relation controls the universality even in the presence of interactions. As the Aubry-Andr\'e model can be derived from a 2D tight-binding Hamiltonian on a square lattice in the presence of a magnetic field, the robustness to interactions (and perhaps other perturbations) of the exponents may originate from the observation that the Diophantine equation can be derived non-perturbatively just considering the properties of the magnetic translation group where $Q_k=\sigma_H$ is the total Hall conductivity \cite{Kraus2012TopologicalEquivalence,Dana_1985}. In fact, in the incommensurate case, the Diophantine equation relates systems with the same hall conductance per length, $\sigma_H/N$.

  The remainder of the work is organized as follows: Sec~\ref{sec:prelim} is devoted to some essential technical information needed for the rest of the paper. Sec~\ref{sec:nonint} focuses on the non-interacting AA model's critical properties. Sec~\ref{sec:discussion} offers an explanation of the observed universal behavior in terms of the Diophantine equation. We then move on to study the interacting model in Sec.~\ref{sec:int}, and we conclude in Sec.~\ref{sec:conclusion}.

  \section{Preliminaries}\label{sec:prelim}
  
  Throughout all of this work, we will be considering periodic or antiperiodic boundary conditions and system sizes determined by the continued fraction expansion for $\beta$, as is typical \cite{Kohmoto1983,Szabo2018,Tang1986}. The continued fraction expansion for $\beta$ can be written as:
\begin{equation}
    \beta =n_0 +\frac{1}{n_1 + \frac{1}{n_2 + \frac{1}{n_3 + ...}}}=[n_0,n_1,n_2,...]
\end{equation}
where, without loss of generality, we set $n_0=0$ as it does not affect $H_{AA}$. Truncating the series at $n_k$ gives a rational approximation to $\beta$ as $\beta \approx M_k / N_k$ for $M_k$ and $N_k$ coprime. The $N\to \infty$ limit is taken by considering only the system sizes $N_k$ in order to be able to satisfy (anti)periodic boundary conditions.

We will say two $\beta$'s have the same {asymptotic} continued fraction expansion if there exists some natural number $k$ such that, for all $i>k$, the $n_i$ appearing in the continued fraction expansion are the same.
 
 To determine the critical behavior, we will compute the following quantities: the (generalized) fidelity susceptibility, and the superfluid fraction. The fidelity susceptibility is a powerful tool for studying quantum phase transitions  (see Ref.~\onlinecite{gu2010} and references therein). With a generalized version, the exponents $z,\nu$ were extracted for the single particle AA model \cite{Wei2019}, and a transition with $\beta\to0$ in a controlled way with $\nu\approx 0.7$ was found at half-filling \cite{Thakurathi2012}. The fidelity susceptibility is defined as
 \begin{equation}
    \chi_F = \lim_{\delta \lambda \to 0} \frac{-2 \ln F}{\delta \lambda^2};\qquad F =| \langle \Psi(\lambda+\delta \lambda) | \Psi(\lambda) \rangle|.
\end{equation}

The superfluid fraction was used by Refs.~\cite{Ray2015,Chaves1997,Szabo2018} on this model. It is given by
\begin{equation}
    \Gamma = N^2 \frac{d^2E}{d\theta^2}
\end{equation}
where $E(\theta)$ is the energy with twisted periodic boundary conditions and is related to the curvature of the lowest band in the single-particle spectrum case \cite{Szabo2018}. 

These two quantities access certain critical exponents in the following way \cite{Szabo2018,gu2010,Continentino1992}
\begin{equation}
\begin{aligned}
    \chi_F(\lambda=\lambda_{\text{max}}) &\sim N^\mu; \\
    \frac{\chi_F(\lambda=\lambda_\text{max}) - \chi_F(\lambda) }{\chi_F(\lambda)} &= f(N^{1/\nu}(\lambda-\lambda_\text{max}) \\
   \Gamma &= N^{2-z}g(N^{1/\nu}(\lambda-\lambda_\text{max})
\end{aligned}
\end{equation}
The exponent $\nu$ has been extracted in the interacting Aubry-Andr\'e model before using a different quantity in Ref.~\onlinecite{Schuster2002}.\footnote{  Note that there is misrepresentation of the fidelity susceptibility in the literature that says $\chi_F \sim N^{2/\nu}$, but this is not correct. This can most easily be seen in the Kitaev Honeycomb model where $\mu \approx 5/2$ and $\nu \approx 1$ \cite{Yang2008}. However, it is quite common that $\mu=2/\nu$, and we will always be able to compute $\nu$ via the universal functions. We find that $\mu\approx 2/\nu$ with the largest deviation occurring for $\beta =\text{``0''}$ (see Fig.~\ref{fig:betato0transition}).}
 
 Through $\Gamma$, $z$ is difficult to determine as it does not have a peak, but we can extract the value of $z$ in the $V=0$ case through the generalized fidelity susceptibility via the following equation \cite{Wei2019,DeGrandi2010}
\begin{equation}
    \chi_{F,2+2r} = \sum_{n\ne 0} \frac{ | \langle \Psi_n | H_I | \Psi_0 \rangle|^2 }{(E_n - E_0)^{2+2r}}
\end{equation}
where $r=0$ is the usual fidelity susceptibility and $H = H_0 + \lambda H_I$. It is known that $\chi_{F,2+2r}\sim N^{\mu+2zr}$ at the critical point \cite{Wei2019,DeGrandi2010}, which provides an efficient means of extracting $\mu$ and $z$. In the free case, for arbitrary fillings, this is a possible computation because only $\sim N^2$ states contribute (see Appendix~A); once we compute $z$ from the generalized fidelity susceptibility, that same value is used to collapse the $\Gamma$ curves onto each other.

\subsection{Boundary conditions}

Since we are interested in the thermodynamic limit, we expect that boundary conditions do not play such an important role. However, we find that the boundary conditions do influence the finite-size scaling collapse. Therefore, we want to make as consistent a choice as possible. The easiest way to continue is not to consider the fermionic Hamiltonian form Eq.~\eqref{eq:H_AA} above but to consider the spin Hamiltonian:
\begin{equation}
\begin{aligned}
    H &= -\sum_{i=1}^{N-1} J(S_i^+S_{i+1}^- + h.c.) -A J (S_N^+S_1^-e^{i\theta} + h.c.)
    \\&+\sum_{i=1}^N V S_i^z S_{i+1}^z + h_i S_i^z
\end{aligned}\label{eq:spinham}
\end{equation}
where $A=1$ corresponds to periodic boundary conditions (PBC) and $A=-1$ corresponds to antiperiodic boundary conditions (ABC).  We have made the twist in the boundary condition $\theta$ explicit. When we map back to the fermionic Hamiltonian via a Jordan-Wigner transformation, we find that
\begin{equation}
\begin{aligned}
    H &= -\sum_{i=1}^{N-1} J(\hat c_i^\dagger \hat c_{i+1} +h.c.) -P_FAJ(c_N^\dagger c_1 e^{i\theta}+h.c.) \\&+ \sum_i h_i \hat n_i +V \hat n_i \hat n_{i+1}
\end{aligned}\label{eq:fermiham}
\end{equation}
where $P_F = (-1)^{N_F}=(-1)^{N} \prod_i(-2S_i^z)$ up to a shift in the chemical potential. We will set $J=1$ from here on in.
This is a number conserving Hamiltonian, so we have the good quantum number $N_F = N_\uparrow$, and we will study it at the filling $n=N_F/N = N_\uparrow/N$.

Within the spin language, the Hamiltonian exhibits spin-flip symmetry, which relates the ground states $\Psi(\lambda/|J|,\text{sgn}(J),A,\phi,N_\uparrow) \leftrightarrow \Psi(\lambda/|J|,\text{sgn}(J),A,\phi+\pi,N-N_\uparrow)$. Since $\text{sgn}(\Gamma)=(-1)^{N_F}$, the data cannot be collapsed well if $N_F$ takes both even and odd values. We fix this by using the set up in Table~\ref{tab:param choice}. Essentially, this guarantees keeping the spin Hamiltonian the same, though using $\phi\in \{0,\pi\}$, being not a generic value of $\phi$, means that the collapse fails in certain cases and other angles need to be tried. 

\begin{table}[]
    \centering
    \begin{tabular}{c||c|c}
         & $N_F$ even & $N_F$ odd \\
         \hline
         \hline
       $N$ odd & $\phi=\pi, P_F=1$ & $\phi=0, P_F=-1$ \\
       \hline
       $N$ even & $\phi=\pi/2, P_F=1$ & $\phi=3\pi/2, P_F=-1$ 
    \end{tabular}
    \caption{Choice of angle, $\phi$, and $P_F$ for a given $N$ and $N_F$ when computing $\Gamma$ and $\chi_{F,2+2r}$.}
    \label{tab:param choice}
\end{table}

The specification in Table~\ref{tab:param choice} means we are looking at a system with ABC in the single-particle spectrum which is equivalent to studying the $J>0$ model with PBC when $N$ is odd (because of the transformation $c_{2n}\to-c_{2n}$).

\section{Non-interacting case $V=0$}
\label{sec:nonint}
We now study how $\Gamma$ and $\chi_{F,2+2r}$ behave in the free case. We consider only $\beta$ with a periodic continued fraction expansion, and, for simplicity, only those of the form $[..., n,n,n,n,...]$. Specifically, we will focus on the following incommensurate ratios. 
\begin{equation}
    \beta_{nm} = [0,m,n,n,n,...] = \frac{1}{m+\beta_{nn}}
\end{equation}
where $\beta_{nn} = (\sqrt{n^2+4}-n)/2$ are the metallic means. We will also consider $\beta = \text{``0''}$ with best rational approximation $1/N$ for all $N$ \cite{Thakurathi2012}.

\subsection{Results}

We are able to use this to reproduce the results \cite{Szabo2018,Hashimoto1992} (see also Ref.~\onlinecite{Wei2019}) for the critical exponents $z$ in the single-particle spectrum. In this case, $z$ only depends on the asymptotic continued fraction expansion. For $\beta \ll 1$, the exponent $z$ is in fact given by \cite{Szabo2018}
\begin{equation}
    z(\beta_{nm}) \approx 1.1662 \frac{\beta_{nn}^{-1}}{\log(\beta_{nn}^{-1})},
\end{equation}
where it is clear that $z\to \infty$ as $\beta \to 0$.

From now on, we consider fillings with an extensive number of particles. We focus on the sector with $N_F = \lfloor \rho N\rceil$ where $\lfloor x \rceil$ rounds $x$ to the nearest integer, and $\rho\in (0,1)$ is the filling fraction. 

\begin{figure*}
    \centering
    \begin{overpic}[width=.4\textwidth]{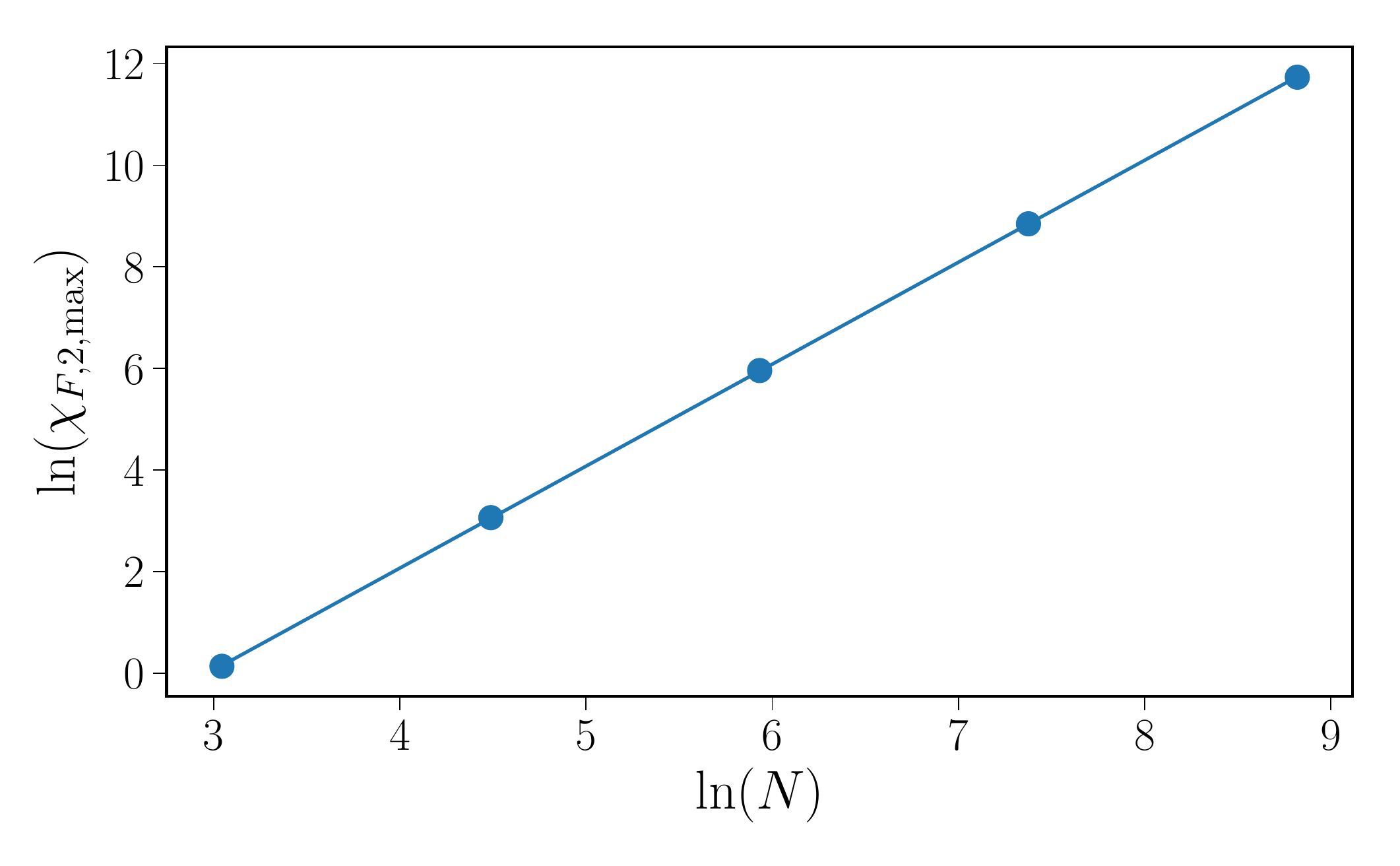}
        \put(0.1,60){$a)$}
    \end{overpic}
    \begin{overpic}[width=.4\textwidth]{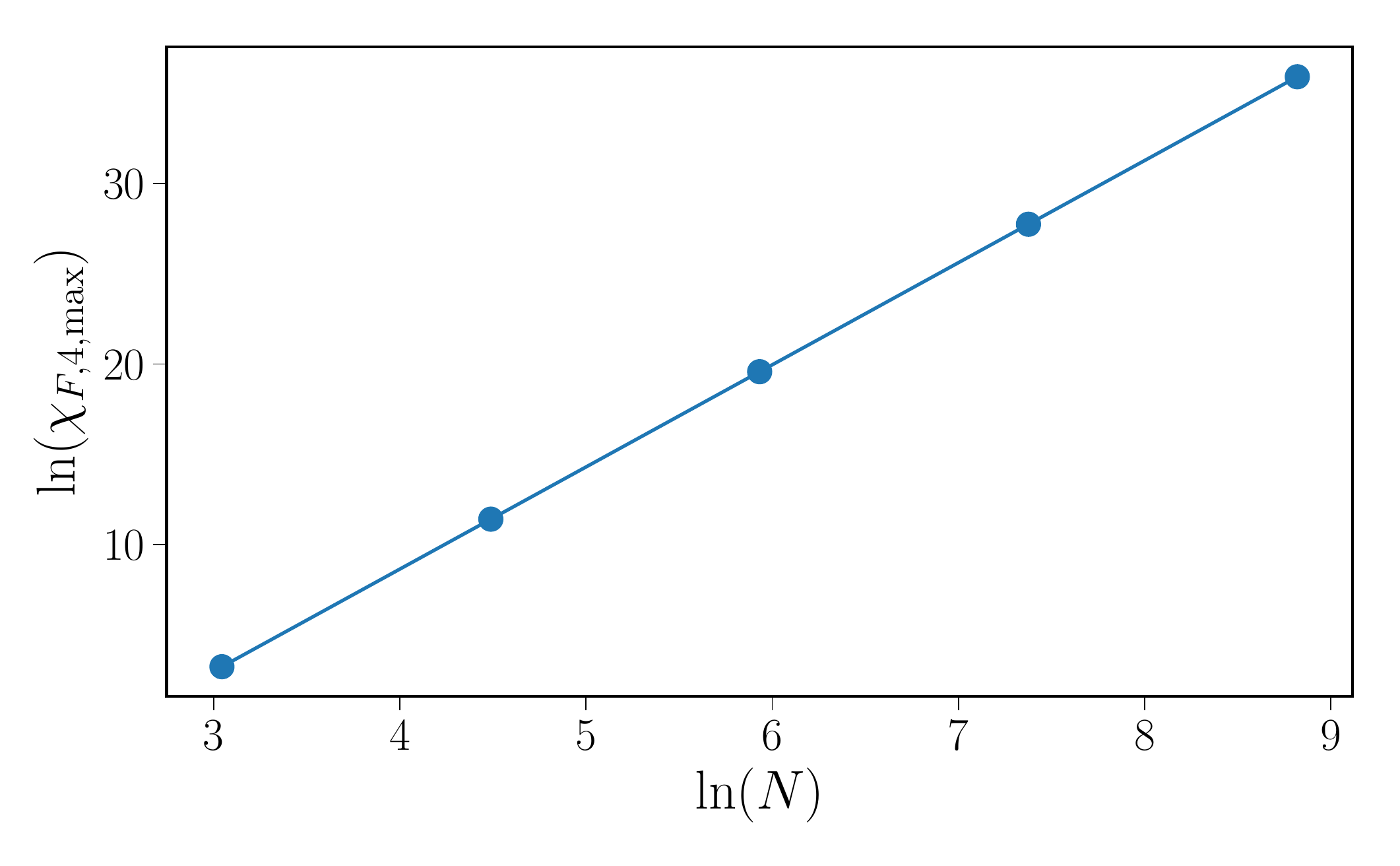}
        \put(0.1,60){$b)$}
    \end{overpic}
    
    \begin{overpic}[width=.4\textwidth]{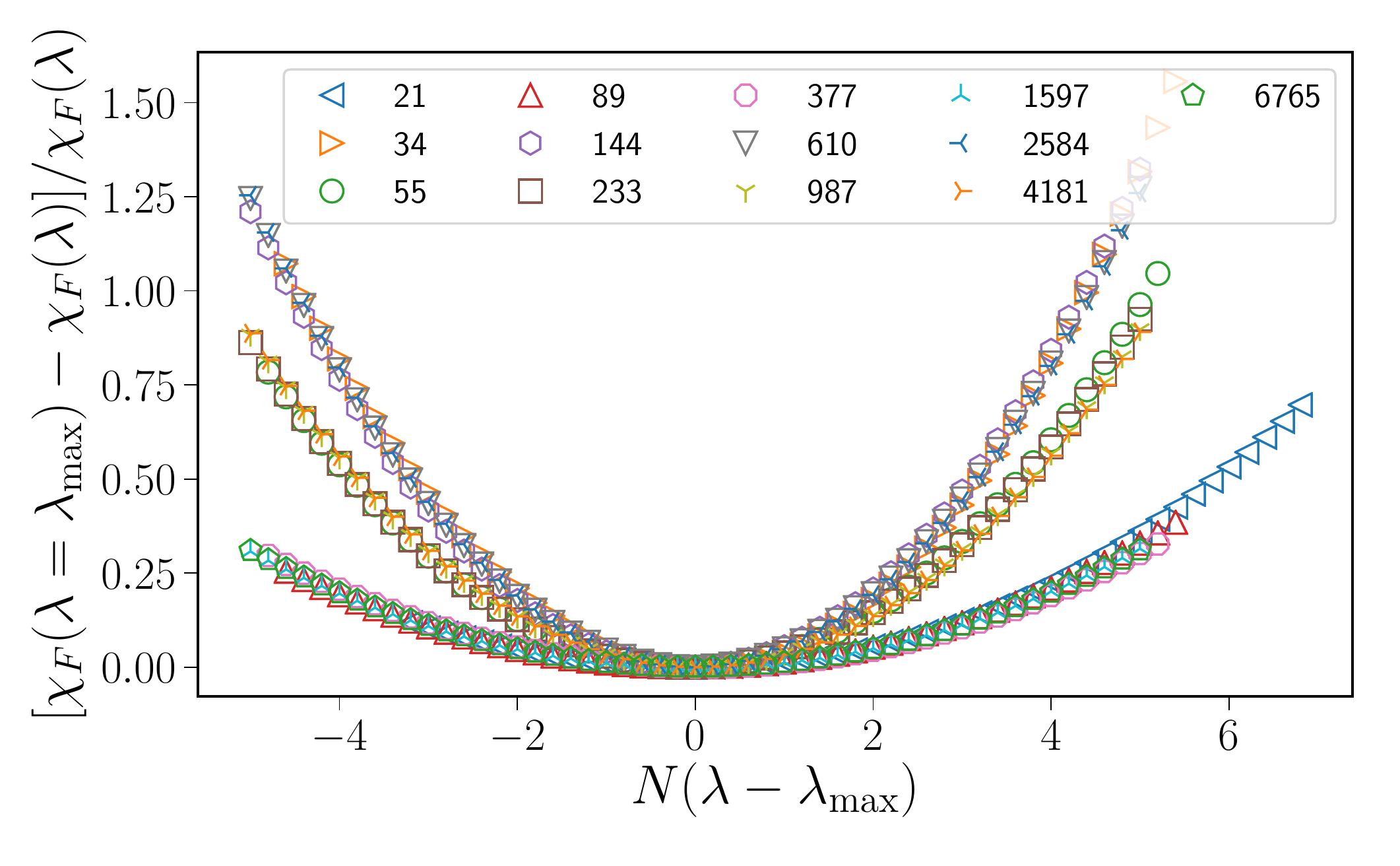}
        \put(0.1,64){$c)$}
    \end{overpic}
    \begin{overpic}[width=.4\textwidth]{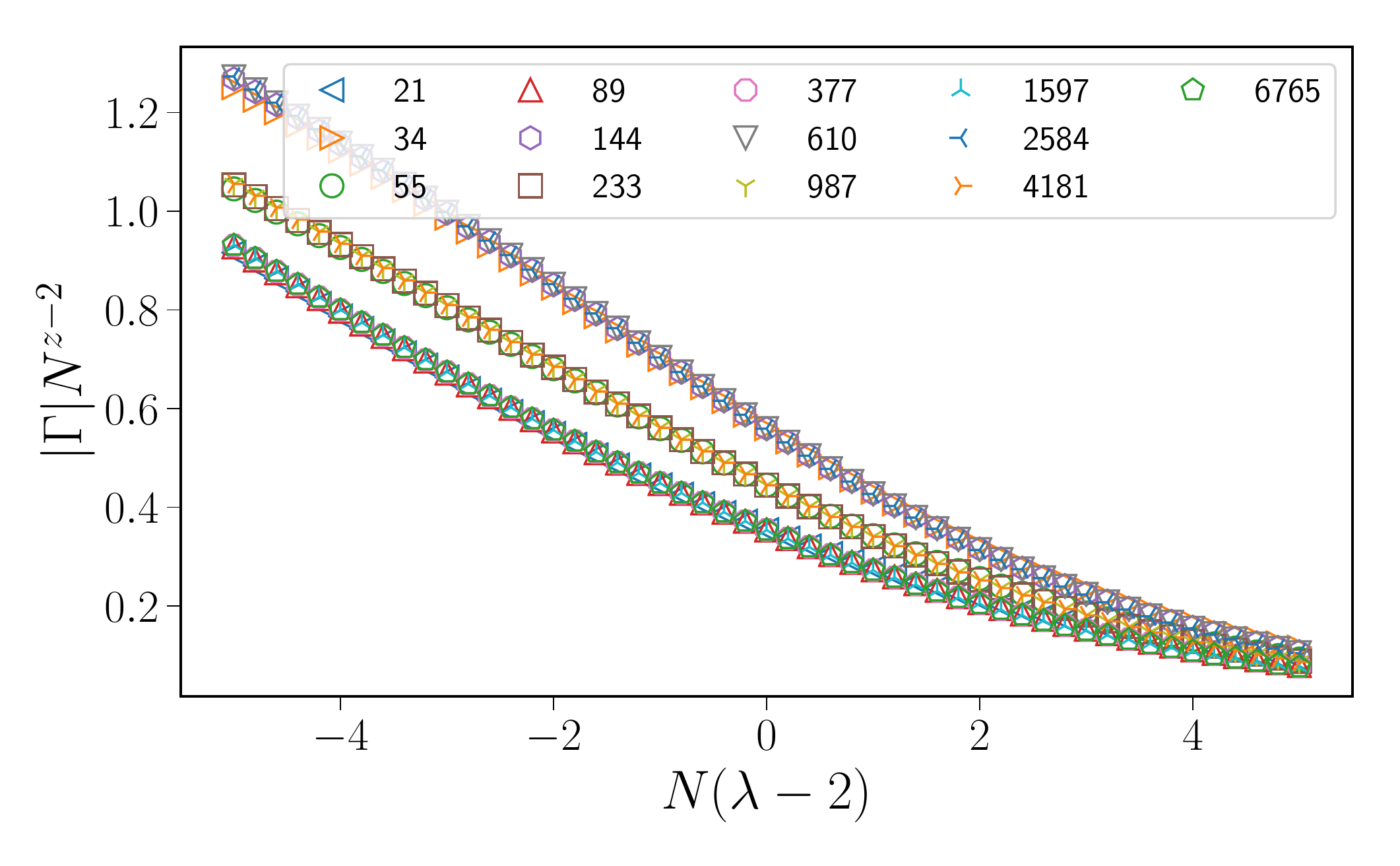}
        \put(0.1,64){$d)$}
    \end{overpic}
    \caption{Scaling of the generalized fidelity susceptibility $\chi_F$ and the superfluid fraction $\Gamma$ for $\beta = \beta_{11}= (\sqrt{5}-1)/2$ at filling $\rho=1/2$ with the angle and $P_F$ specified by Table I. Parts a) and b) show how the maxima of $\chi_{F,2+2r}\sim L^{\mu+2zr}$ from which $\mu=2.0$ and $z=1.8285$ are extracted. In c) and d), the scaling of $\chi_F$ and $\Gamma$, respectively, are consistent with $\mu=2/\nu=2.00$, and $z=1.8285$ collapses all the $\Gamma$ curves onto three universal curves.}
    \label{fig:GR}
\end{figure*}

This investigation leads to a series of observations. First, at half-filling $\rho=1/2$, we notice that not all system sizes are in the same universality class as seen in Fig.~\ref{fig:GR} for the golden ratio $\beta_{11}$. The exponents are the same to a few decimal points, but the scaling functions are different. In this case, the Fibonacci sequence breaks into three subsequences $34, 144, 610, ...$; $21, 55, 233, 987, ...$; and $89, 377, 1597, ...$. This separation into three universality classes has been observed in the exact RSRG scheme \cite{Ostlund1984} and in multi-fractality studies \cite{Tang1986} and the value of the exponent for $\beta_{11}$ agrees with that of \cite{Tang1986, Kohmoto1983,Ostlund1984} for the scaling of the middle part of the spectrum. As a more general pattern, when considering $\beta_{1m}$ for any $m$, we find that they also break into three universality classes with the same exponent $z$.

\begin{figure*}
    \centering
    \begin{overpic}[width=.4\textwidth]{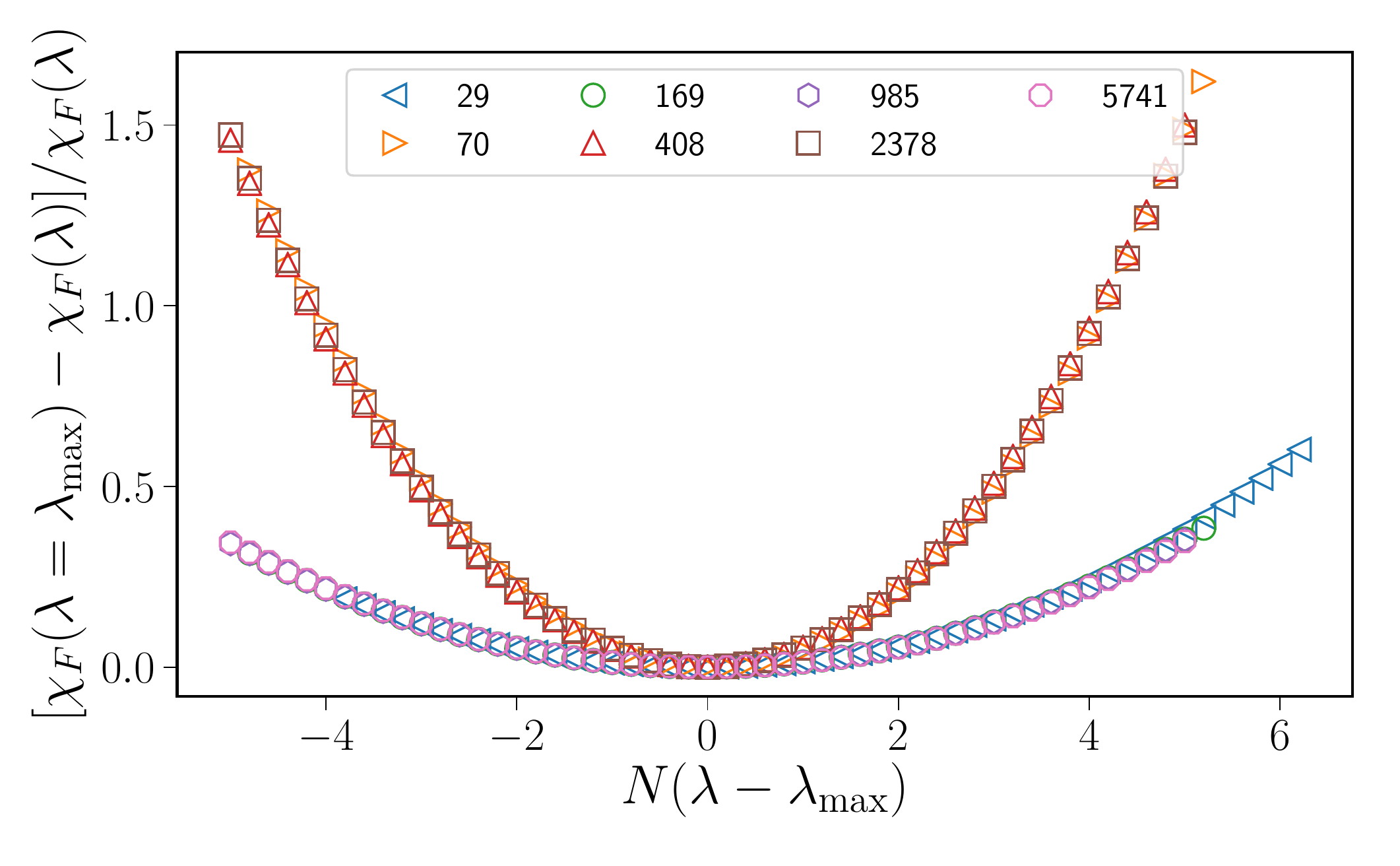}
        \put(0.1,64){$a)$}
    \end{overpic}
    \begin{overpic}[width=.4\textwidth]{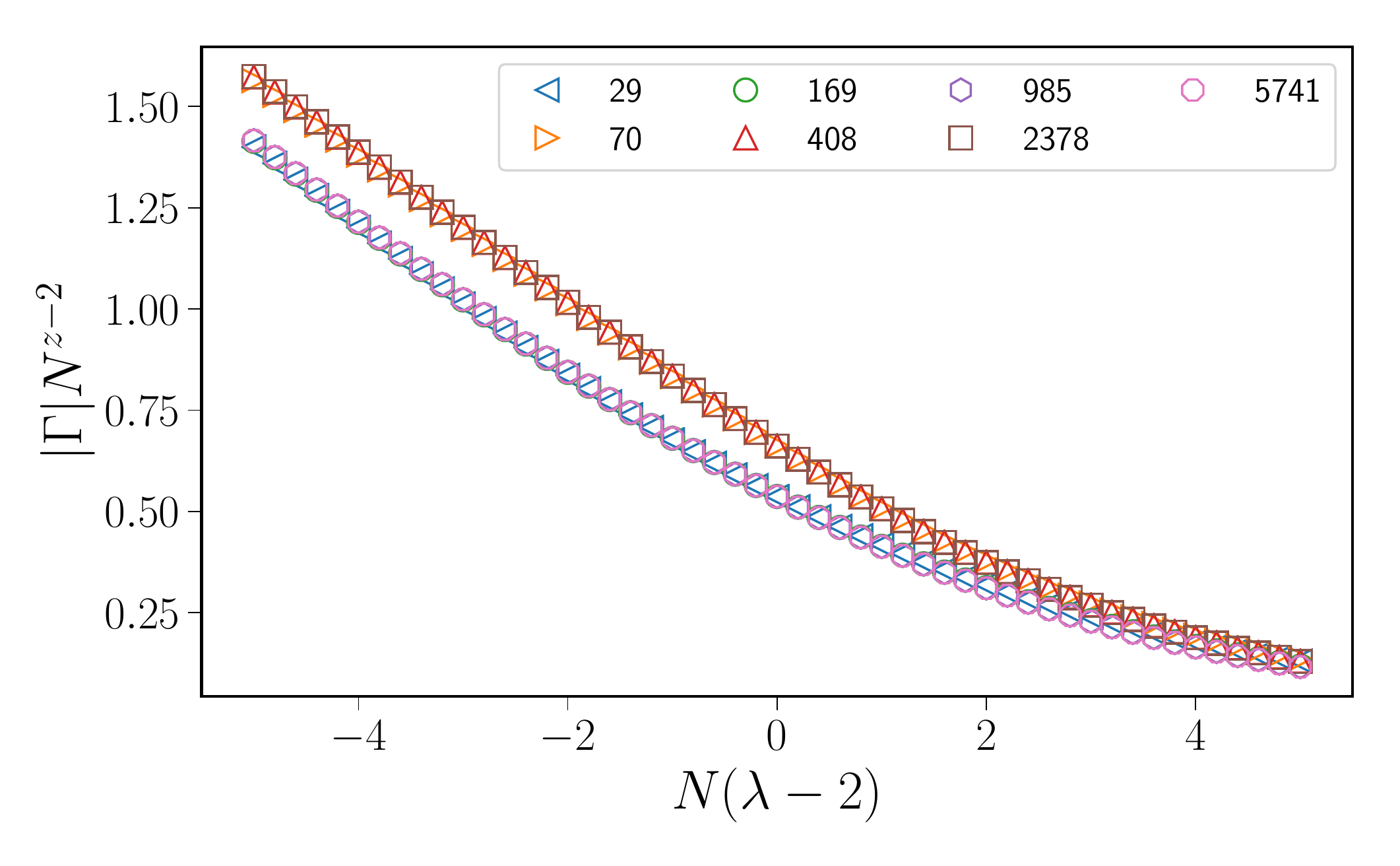}
        \put(0.1,64){$b)$}
    \end{overpic}
    
    \begin{overpic}[width=.4\textwidth]{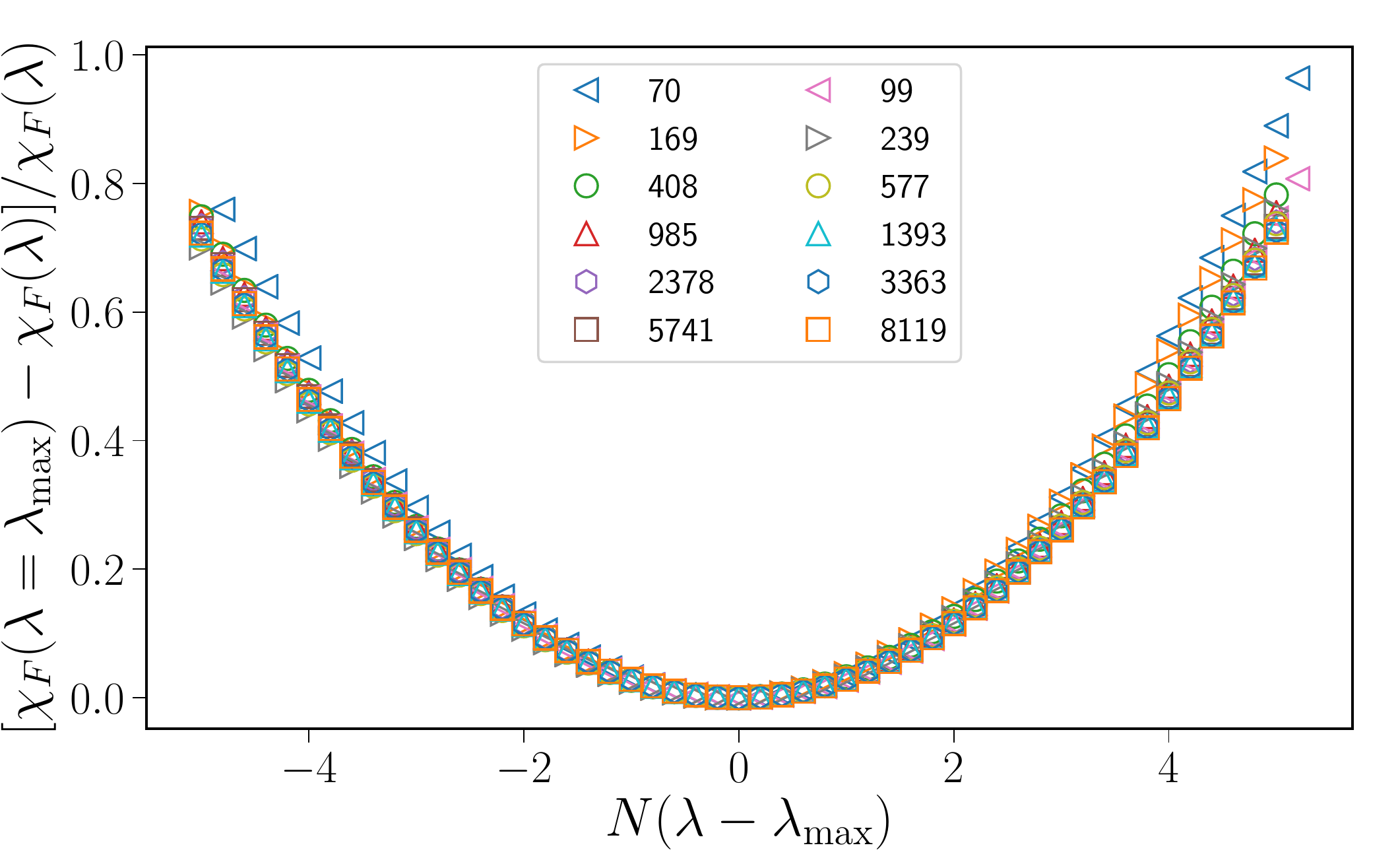}
        \put(0.1,64){$c)$}
    \end{overpic}
    \begin{overpic}[width=.4\textwidth]{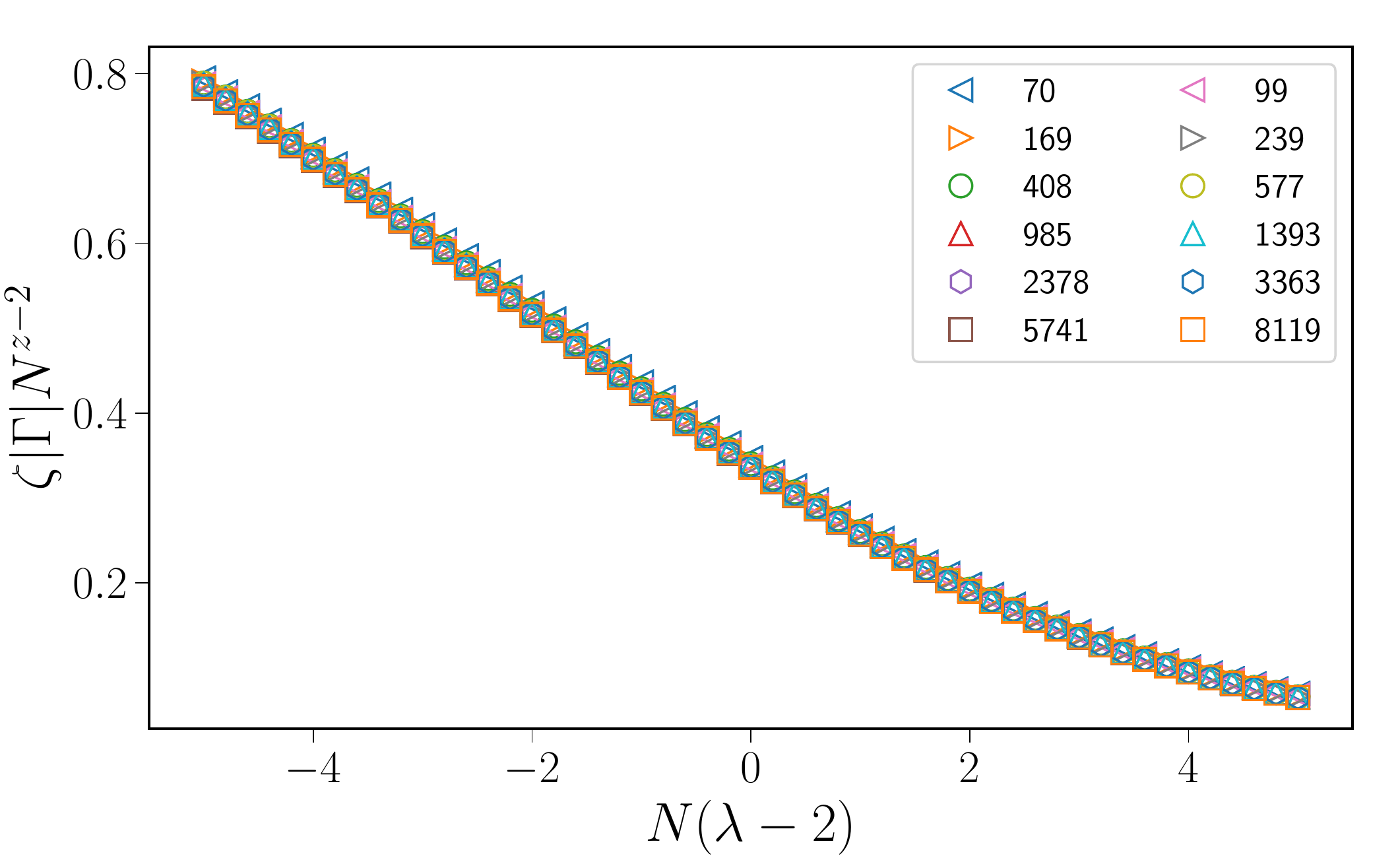}
        \put(0.1,64){$d)$}
    \end{overpic}
    \caption{In a) and b), the scaling collapse of $\chi_F$ and $\Gamma$, respectively, are plotted for $\beta=\beta_{22}$, and there are two universality classes depending on whether $N$ is even or odd. Again, $\mu=2/\nu=2.00$, and the same value of $z=2.0875$ scales all $\Gamma$ curves onto each other. In c) and d), the same quantities are plotted for $\beta = \beta_{21}$  with a filling of $\rho=1/2$ when $N=99,239,577,1983,3363,8119$, and $\beta=\beta_{22}$ with a filling of $\rho=1-1/\sqrt2$ otherwise. Still $\nu=1$, but $z=1.575$. Up to the normalization of $\Gamma$ ($\zeta = 1.0$ for $\beta_{21}$ and $\zeta = 0.6605$ for $\beta_{22}$) the two curves belong to the same universality class but at different fillings. The finite size effects are worse for $\beta_{22}$ because of the irrational filling. However, once $N$ is large enough for the filling to be well approximated, the  collapse is very good. $\phi$ and $P_F$ are set according to Table~\ref{tab:param choice} }
    \label{fig:SR}
\end{figure*}

However, when we now consider the silver-ratio, $\beta_{22}$ and association $\beta_{2m}$ at $\rho=1/2$, we notice that $\beta_{22}$ splits into only two universality classes and $\beta_{21}$ splits into only one, and, between the two $\beta$, $z$ is different (see Fig.~\ref{fig:SR}). Moreover, when $\beta_{22}$ is at a filling of $1-1/\sqrt{2}$, it has the same exponent as $\beta_{21}$ at half-filling and the curves collapse onto each other after a global rescaling suggesting that they belong to the same universality class. We have also checked explicitly that $\beta_{22}$ and $\beta_{21}$ are in the same universality class at $\rho=1/3$. 

\begin{figure*}
    \centering
    \begin{overpic}[width=.33\textwidth]{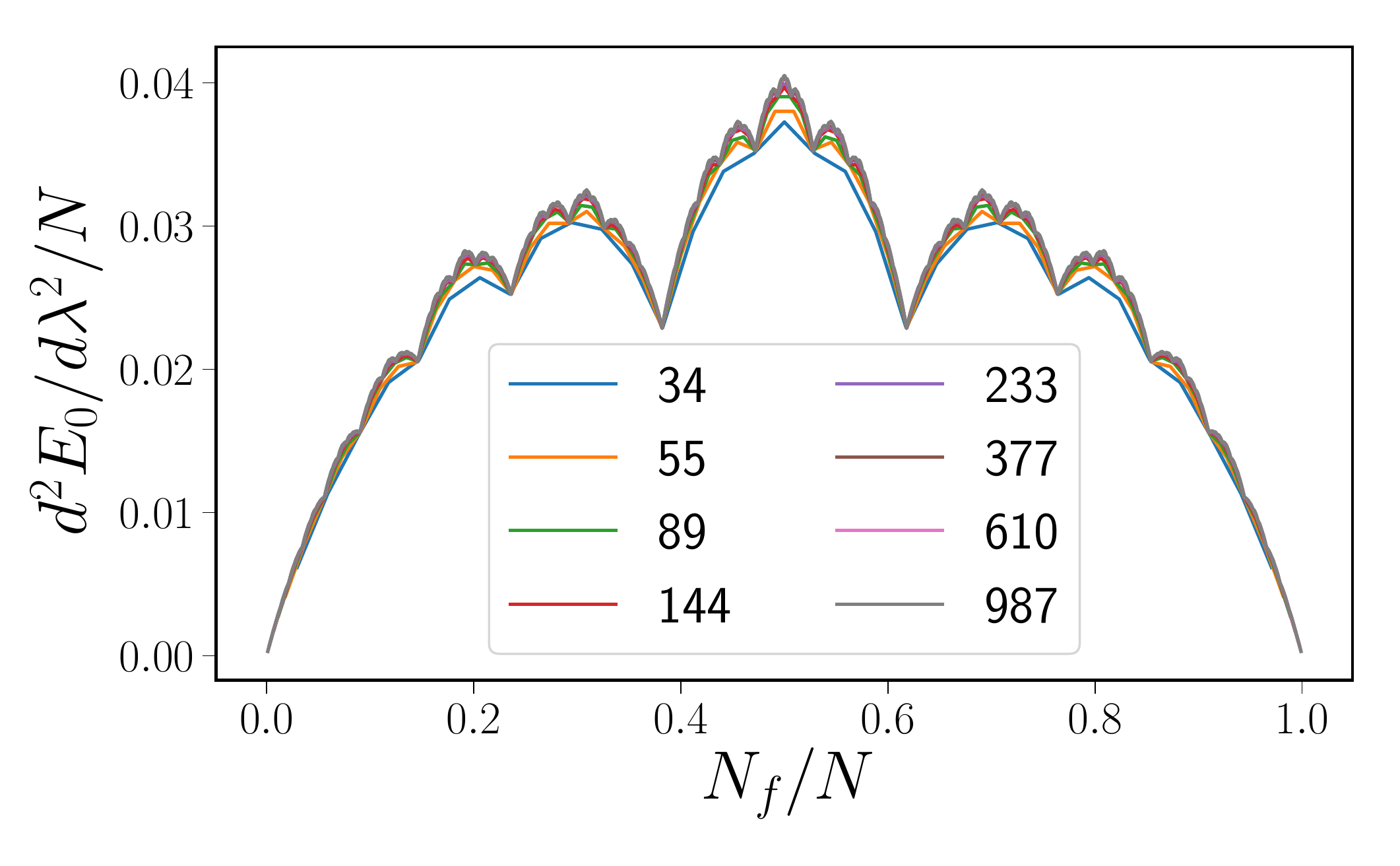}
        \put(0.1,60){$a)$}
    \end{overpic}
    \begin{overpic}[width=.33\textwidth]{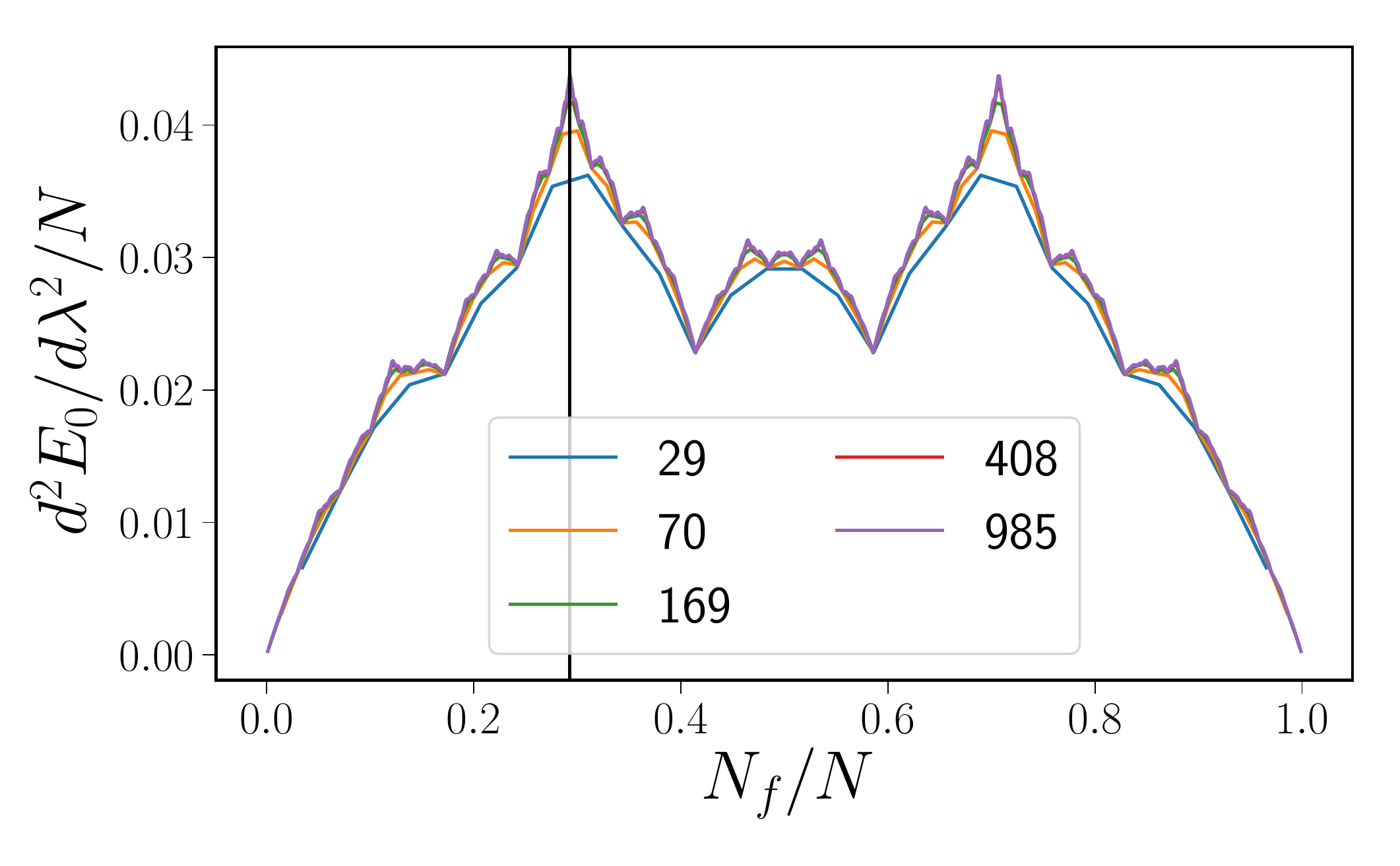}
        \put(0.1,60){$b)$}
    \end{overpic}\begin{overpic}[width=.33\textwidth]{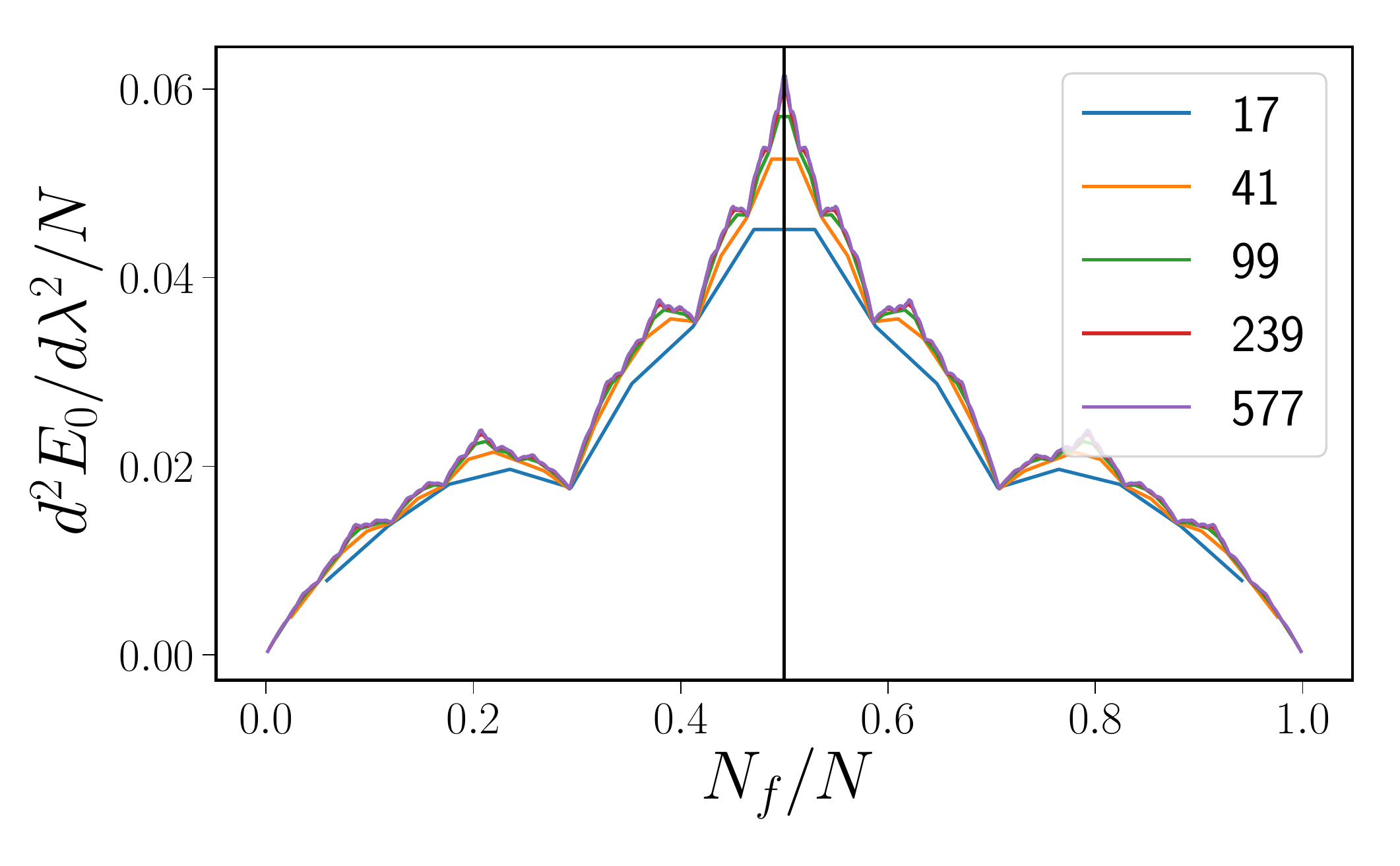}
        \put(0.1,60){$c)$}
    \end{overpic}
    \caption{The second derivative of the ground state energy per particle at $\lambda=2$, $d^2E/d\lambda^2/N$, for a) $\beta_{11}, b) \beta_{22}$, and c) $\beta_{21}$ is plotted against $N_F/N$, the filling fraction. A clear fractal structure emerges. Note the difference between $\beta_{22}$ and $\beta_{21}$ at half-filling, and note the similarity at the filling fraction indicated by the black vertical line at $n=1-1/\sqrt{2}$ for $\beta_{22}$ and $n=1/2$ for $\beta_{21}$ where the same critical exponent is observed (see Fig.~\ref{fig:SR}).}
    \label{fig:d2Efractals}
\end{figure*}

Since the second derivative of the Free energy $\partial^2 E/\partial \lambda^2=\chi_{F,1}$, this quantity also has access to the exponent $z$, so we plot $\chi_{F,1}/N$ v.s. $\rho$ in Fig.~\ref{fig:d2Efractals}. We notice that a fractal shape emerges, which makes it clear that $\rho=1/2$ for $\beta_{21}$ is the same as $\rho = 1-1/\sqrt{2}$ for $\beta_{22}$.

When we broaden our scope to $\beta_{3m}$ and $\beta_{4m}$ (beyond which, the number of accessible system sizes is small) and to filling fractions $1/3$ and $1/4$, we find the exponents in Table~\ref{tab:exponents}. Based on these results, we conjecture that when the filling is $1/q$ and the filling fraction is $\beta_{pm}$, the exponents can be different if the greatest common divisor of $q$ and $p$ is not 1. (see next section~\ref{sec:discussion} for more details and Appendix~D).

\begin{figure*}
    \centering
    \begin{overpic}[width=.4\textwidth]{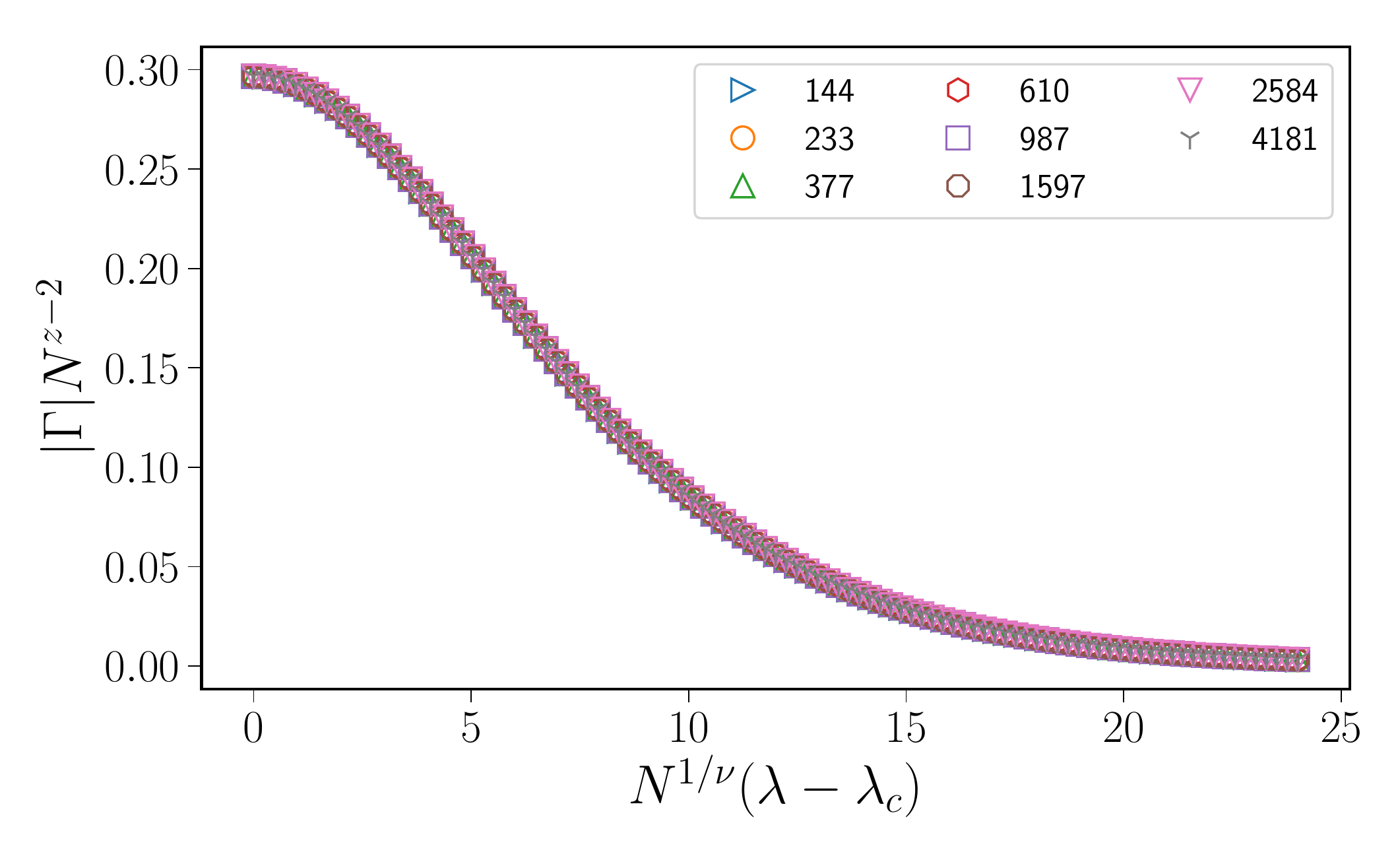}
        \put(0.1,60){$a)$}
    \end{overpic}
    \begin{overpic}[width=.4\textwidth]{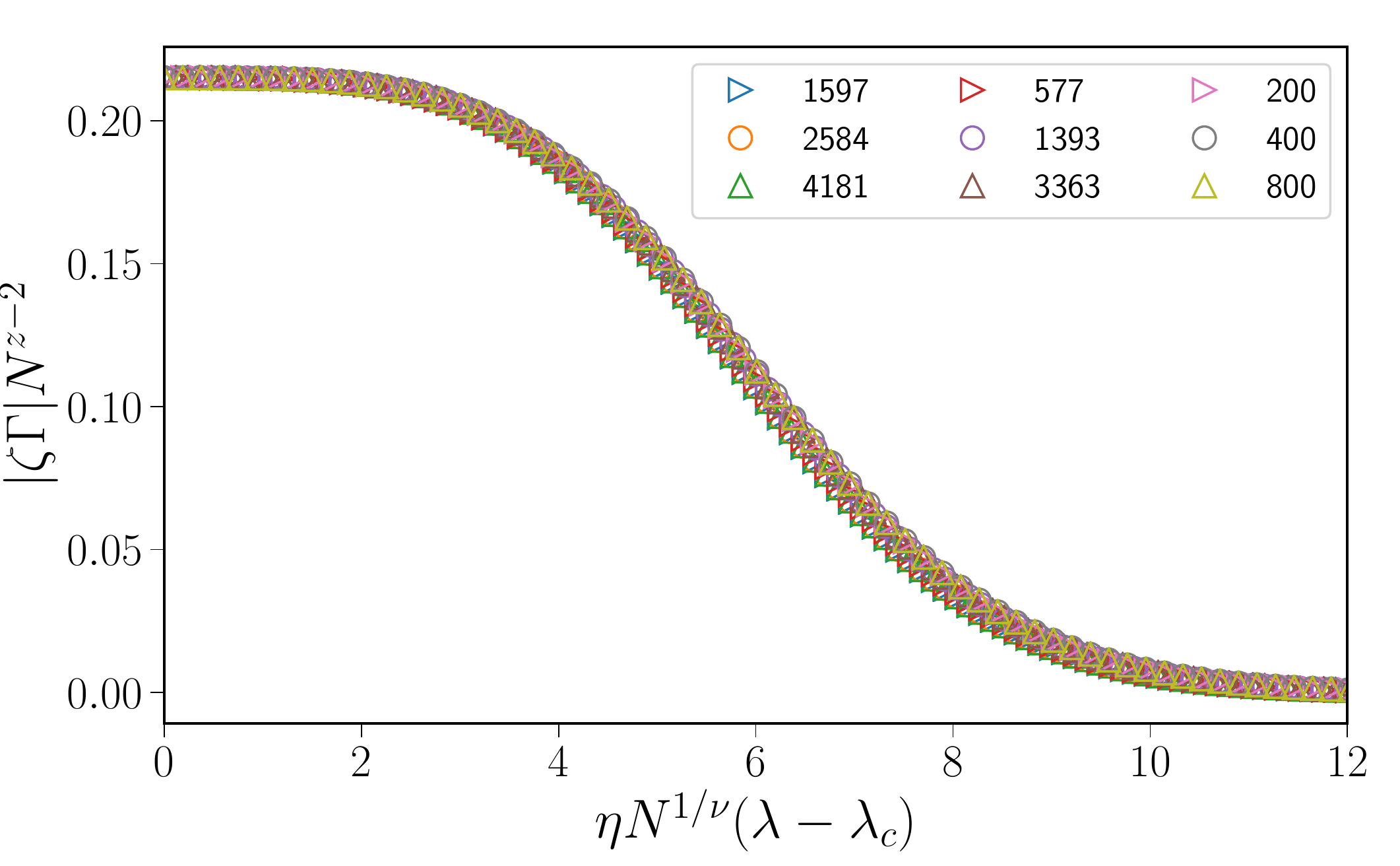}
        \put(0.1,60){$b)$}
    \end{overpic}
    
    \begin{overpic}[width=.4\textwidth]{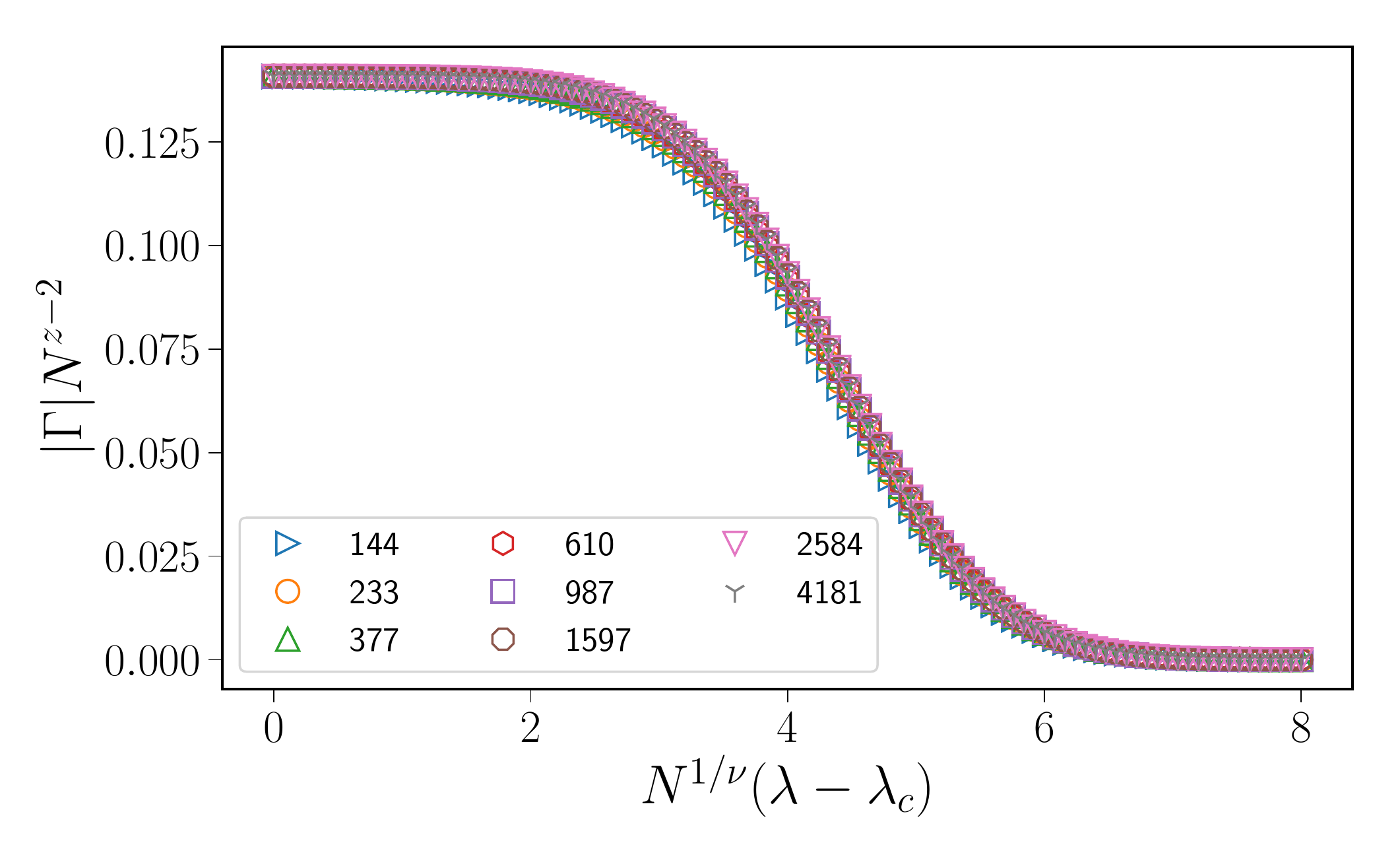}
        \put(0.1,60){$c)$}
    \end{overpic}
    \begin{overpic}[width=.4\textwidth]{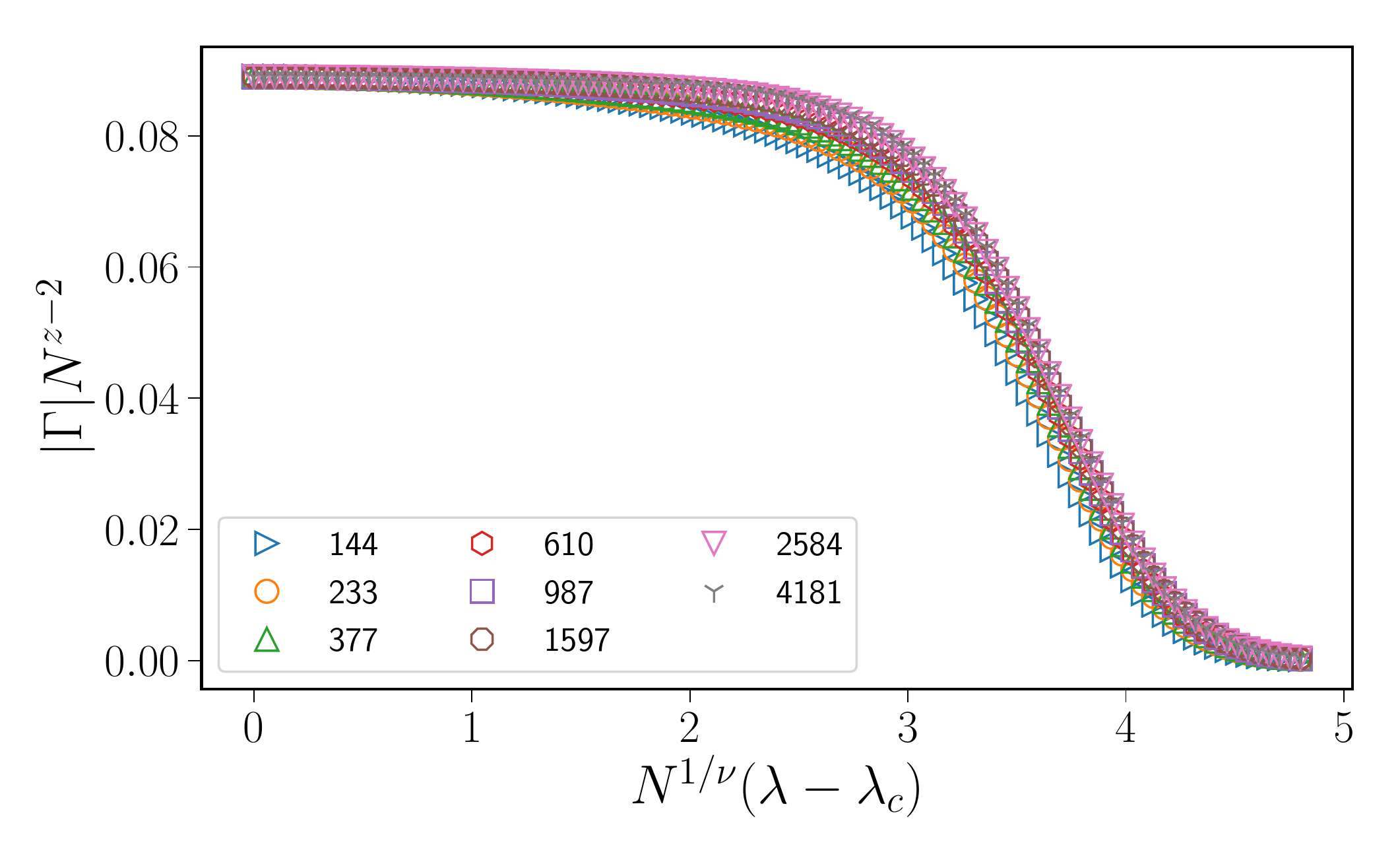}
        \put(0.1,60){$d)$}
    \end{overpic}
    \caption{At commensurate filling, the transition moves from $\lambda_c=2$ to $\lambda_c=0$ because the resonance condition is fulfilled for some fixed $n=\nu$. The dynamic critical exponent is always $z=1$. In a) we have $\beta=\beta_{11}$ and $\rho=1-\beta_{11}$ with $\nu=1$. In b) we plot several $N$ for $\beta=\{\beta_{11}, \beta_{21},1/4\}$ at $\rho=\{2\beta_{11}-1,2\beta_{21}-1,1/2\}$ and we see that they have the same $\nu=2$ and $z=1$. The curves collapse onto each other when $\zeta = \{1,0.7007,0.6755\}$ and $\eta=\{1,1,1.565\}$. In c) we have $\beta=\beta_{11}$ and $\rho=3\beta_{11}-2$ with $\nu=3$. In d) we have $\beta=\beta_{11}$ and $\rho=5\beta_{11}-3$ with $\nu=5$. Note that the finite-size effects get worse as $\rho\to 0$. }
    \label{fig:Commfillfree}
\end{figure*}

Motivated by Ref.~\onlinecite{Schuster2002}, we also consider commensurate fillings $\rho=n\phi-m$ for $n=1,2,3,5$ and $m$ chosen so that $\rho \in (0,1)$. In this case, the transition occurs not at $\lambda=2$, but at $\lambda=0$. That is, a gap immediately opens up because of the close relationship between the Fermi momentum $k_F=\pi\rho$ and the perturbation at $k=2\pi\beta$. The results are shown in Fig.~\ref{fig:Commfillfree}. For all $n$, we consider $\phi=\beta_{11}$ and find that $\nu=n$ and $z=1$ as explained by the perturbation theory analysis in Ref.~\onlinecite{Thakurathi2012}. For $n=2$, we additionally show $\beta=\beta_{12}$ and $\beta=1/4$ at the corresponding commensurate fillings and see that they collapse together onto the same curve.

\begin{table*}[]
    \centering
    \begin{tabular}{c||c|c|c|c}
        $\beta$ & $(z,p), n=1/2$ & $(z,p), n=1/3$ & $(z,p), n=1/4$ & other $(n,z,p)$  \\
        \hline
        \hline
         $\beta_{11}=1-\beta_{12},\beta_{14},\beta_{13}$ & (1.8285,3) & (2.00,4)  & (2.0,6) & \\
         $\beta_{22}$ & (2.0875,2) & ($1.97\pm 0.01$,4) &($-$,8) & ($1-1/\sqrt{2}$,1.575,1) \\
         $\beta_{21}=1-\beta_{23},\beta_{25}$ & (1.575,1) & ($1.97\pm 0.01$,4) & (2.09,2) & \\
         $\beta_{24}$ & (2.0875,2) & ($1.97\pm 0.01$,4) & ($-$,8) & \\
         $\beta_{33}$ & (2.0,3), & (2.24,2) & ($-$,6) & (1/(3$\beta_{31}$),1.67,1)\\
        $\beta_{32}$ & (2.0,3) & (2.02,1) & ($-$,6) & ($\beta_{32}/(3\beta_{31}$),1.67,1) \\
        $\beta_{31}$ & (2.0,3) &(1.67,1) & ($-$,6) \\
       $\beta_{44}$ & (2.374,2) & ($-$,4) & ($-$,2)\\ 
         $\beta_{43}$ & (1.518,1) & ($-$,4)& (2.57,1) \\
        $\beta_{41}$ & (1.518,1) & ($-$,4) & (1.815,1) & \\

    \end{tabular}
    \caption{The exponents extracted from finite-size scaling at various fillings. The number of universality classes is determined by scaling collapse as in Fig.~\ref{fig:GR} and Fig.~\ref{fig:SR} and via the Diophantine equation where $p$ is determined by the period of the repeating values of $|Q_k|/N_k$ for large $k$ for $\beta_{mn}$. In the cases where $z$ and $p$ agree, the same sequence of $|Q_k|/N_k$ appears suggesting the Diophantine equation, Eq.~\eqref{eq:diophantineInText}, probes the universal properties. Indeed, the Diophantine equation predicts which $\beta$ and $n$ to consider to fill the last column, and predicts why differences only occur in the first column when $\beta_{mn}$ has $m$ even. 
    If no value of $z$ is reported, it is because a good collapse was not seen for system sizes $N<10^4$; in those cases, $p$ was determined via the Diophantine equation alone. It is worth noting that a value $z\approx2$ is the expected value for a ``generic'' filling as that is near the peak of the distributions of $1/\alpha$'s in the multifractal analysis \cite{Szabo2018, Tang1986}.
    For $\beta_{2n}$, an error bar of $\pm 0.01$ is given because the large value of $p$ means only two curves fell into each universality class for $N<10^4$; yet, the finite-size scaling is very sensitive to $z$ due to the large system sizes, and the collapse does not work well beyond the reported range.} 
    \label{tab:exponents}
\end{table*}

\begin{figure*}
    \centering
    \begin{overpic}[width=.40\textwidth]{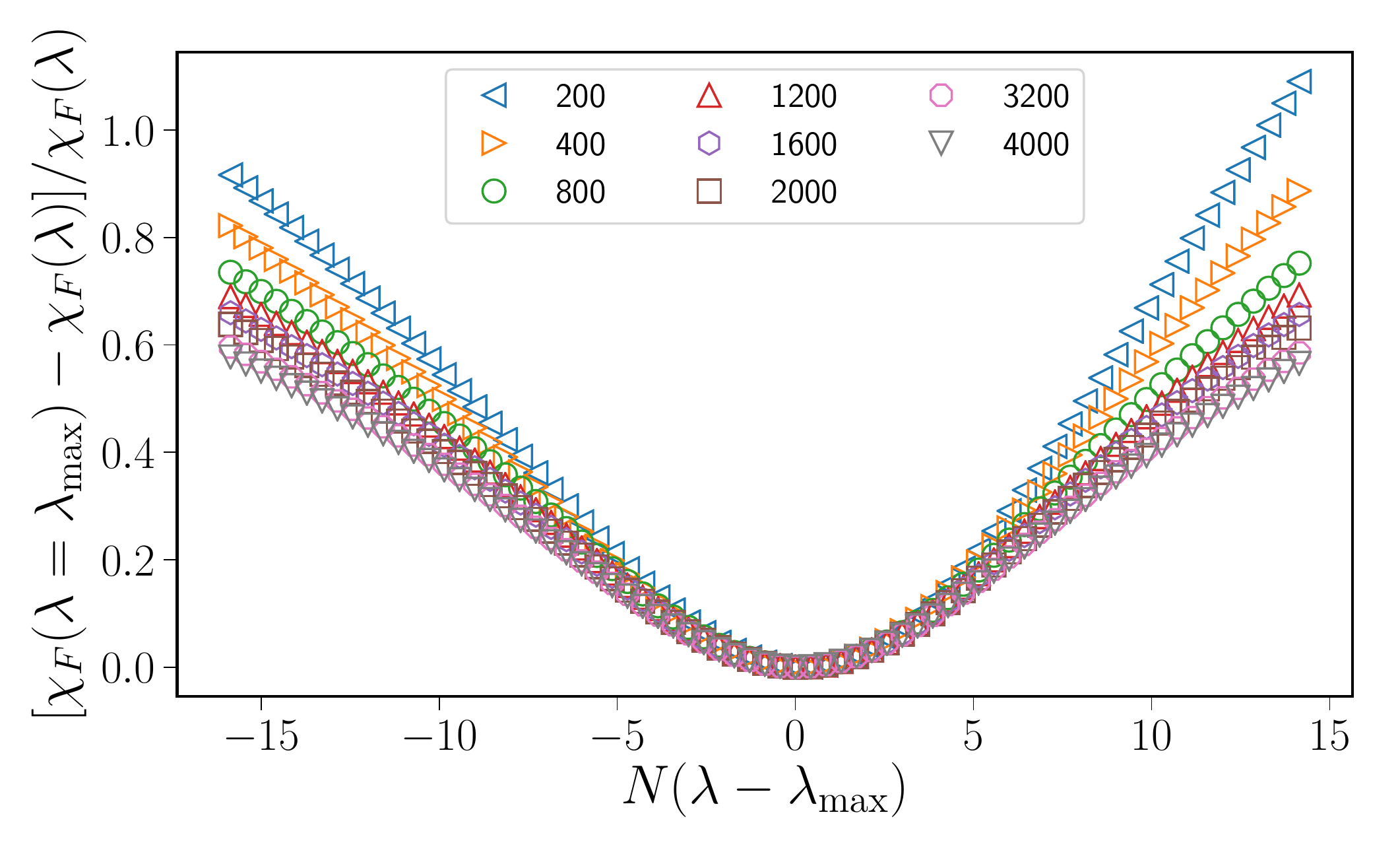}
        \put(0.1,64){$a)$}
    \end{overpic}
    \begin{overpic}[width=.40\textwidth]{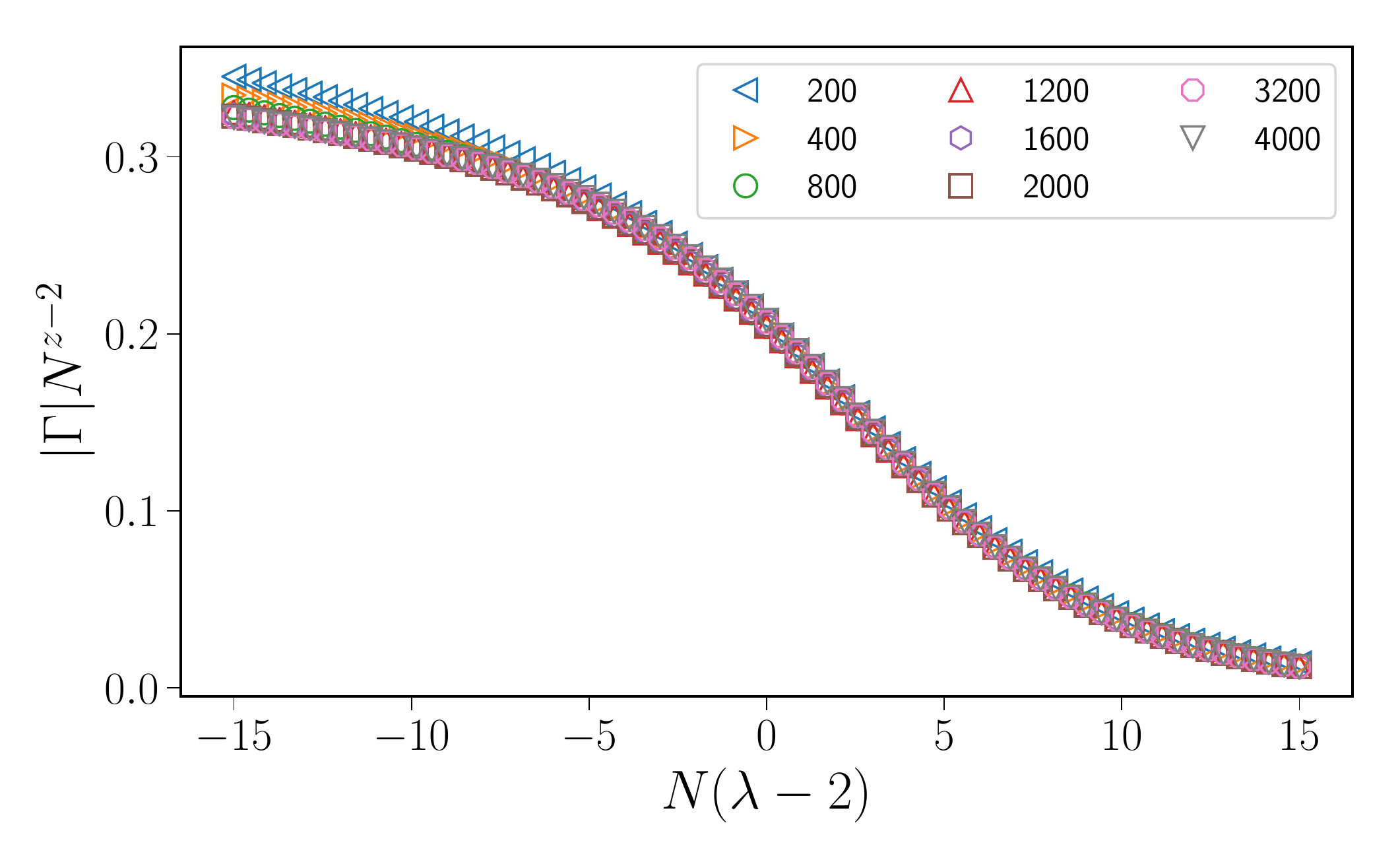}
        \put(0.1,64){$b)$}
    \end{overpic}
    \caption{For $\beta = 1/N$ at filling $N_F=N/2$, we plot the scaling quantities $\chi_F$ and $\Gamma$ in a) and b) respectively. We extract exponents $\nu=1.0$ and $z=1.245$ and $\lambda_c=2.0$. The fidelity susceptibility does not collapse well far from the transition, but $\Gamma$ does, which is why Ref.~\onlinecite{Thakurathi2012} underestimated $\nu \approx 0.7$ as the scaling of the width of the curve. When we fit the maximum of $\chi_F\sim L^\mu \log(L)$ we find $\mu=2.1 \approx 2/\nu$ as opposed to the value of $\mu=2.25$ from Ref.~\onlinecite{Thakurathi2012} where such a logarithmic correction has not been not included. }
    \label{fig:betato0transition}
\end{figure*}

\subsection{Discussion \label{sec:discussion}}

In the case that the filling is commensurate, there is a resonance at $n$th order in  perturbation theory because $\rho = n\beta -m$, and $n$ determines the universality class of the transition as $\nu=n$ \cite{Thakurathi2012}.  When the filling is not commensurate, we can still consider the same equation and resulting Diophantine equation, and it is known that for $\lambda \ll 1$, the most important terms in perturbation theory are those with large $n$ that nearly satisfy the Diophantine equation \cite{Mastropietro2015Gaps}. In the opposite limit, when $\lambda \gg 1$, if we consider when the dominant component of the single-particle wave function changes as we tune $\phi$, we get the same Diophantine equation \cite{TKNN}, where $n$ is the Chern number \cite{ni2019,Kraus2012TopologicalEquivalence}.

We conjecture that the Diophantine equation, at incommensurate fillings, determines the dynamic critical exponent $z$ by controlling how the energy gap vanishes at the $\lambda=2$ transition. With this conjecture, we can understand all the above observations. Noting that we always approximate $\beta\approx \beta_k=M_k/N_k$ and $N_F=\lfloor \rho N_k\rceil$, the resulting Diophantine equation is
\begin{equation}\label{eq:diophantineInText}
    M_k Q_k - P_k N_k = \pm N_F.
\end{equation}
where we restrict $|Q_k|\le N_k/2$ (as the resonance condition is satisfied at the lowest value of $Q_k$) and, of the two possible values of the RHS, we pick the one that gives the smallest $|Q_k|$. We conjecture that $|Q_k|/N_k$ for $k\gg 1$ determines the universality class, and the quantity has a period of $p$ values corresponding to the $p$ universal functions observed (e.g. $p=3$ for $\beta_{11}$ and $p=2$ for $\beta_{22}$ at half-filling). 

This Diophantine equation can be solved exactly with knowledge of the continued fraction expansion of $\beta=[0,n_1,n_2,...]$. First, we solve the case where $\pm N_F=\pm 1$, which is given by $(Q_k',P_k')=\pm(N_{k-1},M_{k-1})$ if $n_{k-1}\ne 1$ and $(Q_k',P_k')=\pm(N_{k-2},M_{k-2})$ otherwise. Then, the solution to the original equation is given by $Q_k = Q_k'N_F \text{ mod } N_k$ and $P_k = P_k' N_F \text{ mod } M_k$.

Immediately, in the single particle spectrum, we easily compute that $|Q_k|/N_k = N_{k-1}/N_k$ (or $N_{k-2}/N_k$). It suffices to show, then, that $N_{k-1}/N_k$ just depends on the asymptotic part of the continued fraction expansion.  This can easily be shown in the case of $\beta=\beta_{nm}$ since (denoting $\beta_k=M_{\beta,k}/N_{\beta,k}$)
\begin{equation}
    \frac{M_{\beta_{nm},k}}{N_{\beta_{nm},k}} = \frac{N_{\beta_{nn},k}}{N_{\beta_{nn},k}m+M_{\beta_{nn},k}}
\end{equation}
Note that, as ${M_{\beta_{nn},k}}/{N_{\beta_{nn},k}}$ is a reduced fraction, ${M_{\beta_{nm},k}}/{N_{\beta_{nm},k}}$ is as well. Therefore,
\begin{equation}
    \lim_{k\to\infty}N_{\beta_{nm},k}/N_{\beta_{nm},k-1}  =N_{\beta_{nn},k}/N_{\beta_{nn},k-1} \approx n +\beta_{nn},
\end{equation}
where we used $M_{\beta_{nn},k}/N_{\beta_{nn},k}\to\beta_{nn}$ as $k\to\infty$. This argument can be easily extended to the general case, so $p=1$ and $|Q_k|/N_k$ is determined solely by the asymptotic continued fraction expansion consistent with the RSRG scheme \cite{Thouless1983,suslov1982localization,Szabo2018}.

Outside of the single-particle spectrum, we worked an explicit example in the introduction that showed $\beta_{21}$ and $\beta_{22}$ are predicted not to be in the same universality class at half-filling. Additionally, we can explain why the fractal shape in Fig.~\ref{fig:d2Efractals} appears. In Appendix~B, we use the Diophantine equation and our conjecture to derive that the universality class is the same at a density of $\rho$ and a density of $\rho/\beta_{nn}$ for $\beta_{nn}$. This fact would reproduce a fractal shape as $\rho$, $\rho/\beta_{nn}^k$, and $\rho \beta_{nn}^k$ will all have the same $z$ for any integer $k$. 

Notably, this does not hold for non-metallic means where, for instance, using $\rho=1/2$ and $\rho=1/\sqrt{2}-1/2$ can be shown to be related with the Diophantine equation trivially, which is also seen as the second largest peak within the fractal structure in Fig.~\ref{fig:d2Efractals}c. We have explicitly checked that $\beta_{21}$ at this filling not only has the same $z$ but the universal function controlling $\Gamma$ is the same up to a numerical prefactor.   

Furthermore, we can consider $\beta_{nm}$ and, in a way that can be made rigorous as in the calculation of Appendix~A, we can see
\begin{equation}\label{eq:beta21beta22univ}
    Q\beta_{nm} -P = \rho \iff (Q-Pm)-P\beta_{nn} = \frac{\rho}{\beta_{nm}}.
\end{equation}
Using the above result, we see $\beta_{nm}$ at a filling of $\rho$ should have the same exponent as $\beta_{nn}$ at a filling of $\rho'=\rho/\beta_{nm}$. This observation predicts the relationship between $\beta_{21}$ and $\beta_{22}$ and half-filling and $\rho=(1-1/\sqrt{2})$ filling, respectively, as particle-hole symmetry relates a filling of $\rho$ and $1-\rho$ [see Fig.~\ref{fig:SR} c) and d)].

Finally, consider the similar explicit calculation here for $n$ odd:
\begin{equation}
\begin{aligned}
    0&=Q\frac{1}{m+\beta_{nn}} +P \pm \frac12 \iff \\ 
    0 &= (Q+Pm) + P\beta_{nn} \pm \frac12(m+\beta_{nn})\\
    &=\begin{cases} (Q+Pm \pm \frac{m-n}2) + P\beta_{nn}\pm \frac{1}{2\beta_{nn}} & \text{if $m$ is odd}\\
    (Q+Pm\pm \frac{m}{2}) + P\beta_{nn} \pm \frac{\beta_{nn}}{2} & \text{if $m$ is even}\end{cases}\\
    &=\begin{cases} P' + Q'\beta_{nn}\pm \frac{1}{2\beta_{nn}} & \text{if $m$ is odd}\\
    P' + Q'\beta_{nn} \pm \frac{\beta_{nn}}{2} & \text{if $m$ is even}\end{cases}\\
\end{aligned}
\label{eq:beta_nn_fill}
\end{equation}
In the two cases above, if $n$ is odd, we have been able to absorb an integer into the definition of $P'$ to get rid of the dependence on $m$. Since the critical properties of $\beta_{nn}$ at $\beta_{nn}/2$ and $1/(2\beta_{nn})$ are the same as those at $\rho=1/2$, then all the $\beta_{nm}$ have the same exponents at half-filling if $n$ is odd.

This breaks down if $n$ is even because the $m$ odd case does not give an integer value of $P'$. Generically, we expect that if $\rho=1/q$, $\beta=\beta_{nn}$ and $n$ and $q$ are coprime, then all of the $\beta_{nm}$ will be in the same universality class at filling $\rho=1/q$.  
However, if $n,q$ share a common factor, there will be separate classes. If we consider a filling of $1/3$ this would allow for $\beta_{3m}$ to split into three separate universality classes based on the residual of $m \text{ mod $3$}$. We indeed observe this numerically for the systems we can access. 

To summarize, the Diophantine equation can predict the fractal structure of Fig.~\ref{fig:d2Efractals}, explains the number of universal curves, $p$, predicts which fillings and which $\beta$ belong to the same universal classes, and, in commensurate filling, specifies the exponent $\nu$ directly.

\section{Interacting case $V\ne 0$}\label{sec:int}

The Diophantine equation description of the universality seems particularly pathological, so we check whether it persists in the presence of the simplest form of interactions as that is the most interesting perturbation. Trivially, it will persist with a shift in the chemical potential, but $p$-wave pairing terms would  destroy it because well-defined fermion number is necessary for the Diophantine equation. Another possible addition would be to consider farther neighbor hopping which, however, goes beyond the scope of this work.

In order to study the interacting model, we use the density-matrix renormalization group (DMRG) \cite{White} on Eq.~\eqref{eq:spinham} with $V\ne 0$ as implemented by the iTensor library \cite{iTensor}. We must have PBC or ABC to compute $\Gamma$ to extract $z$. 
This choice of boundary condition makes convergence in the matrix prodcut state (MPS) bond dimension slower, as a truncation error comparable to one that is achieved by bond dimension $m$ in open boundary conditions, requires $m^2$ in periodic boundary conditions~\cite{Schollwoeck}.

\begin{figure*}
    \centering
    \begin{overpic}[width=.33\textwidth]{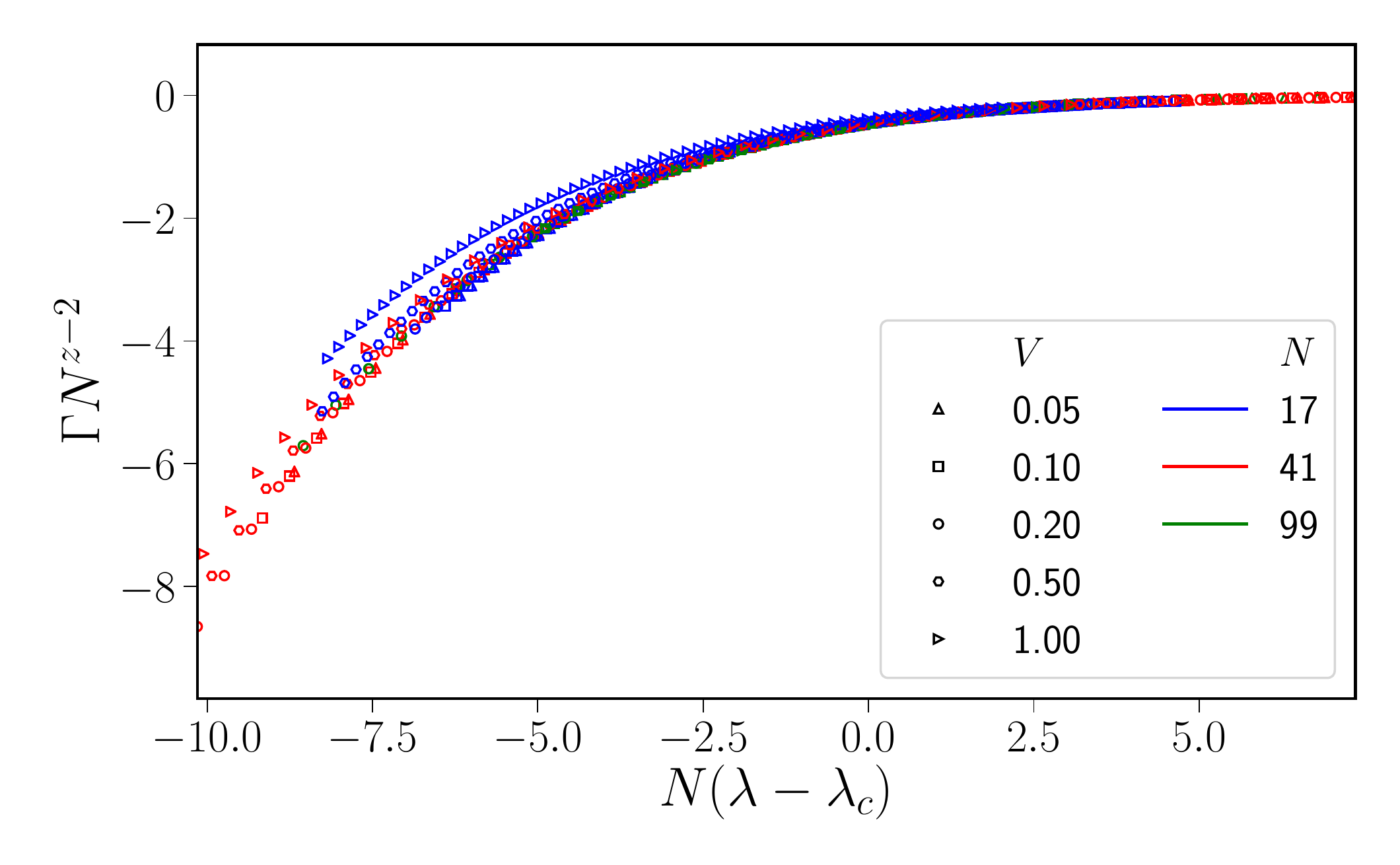}
        \put(0.1,60){$a)$}
    \end{overpic}
    \begin{overpic}[width=.33\textwidth]{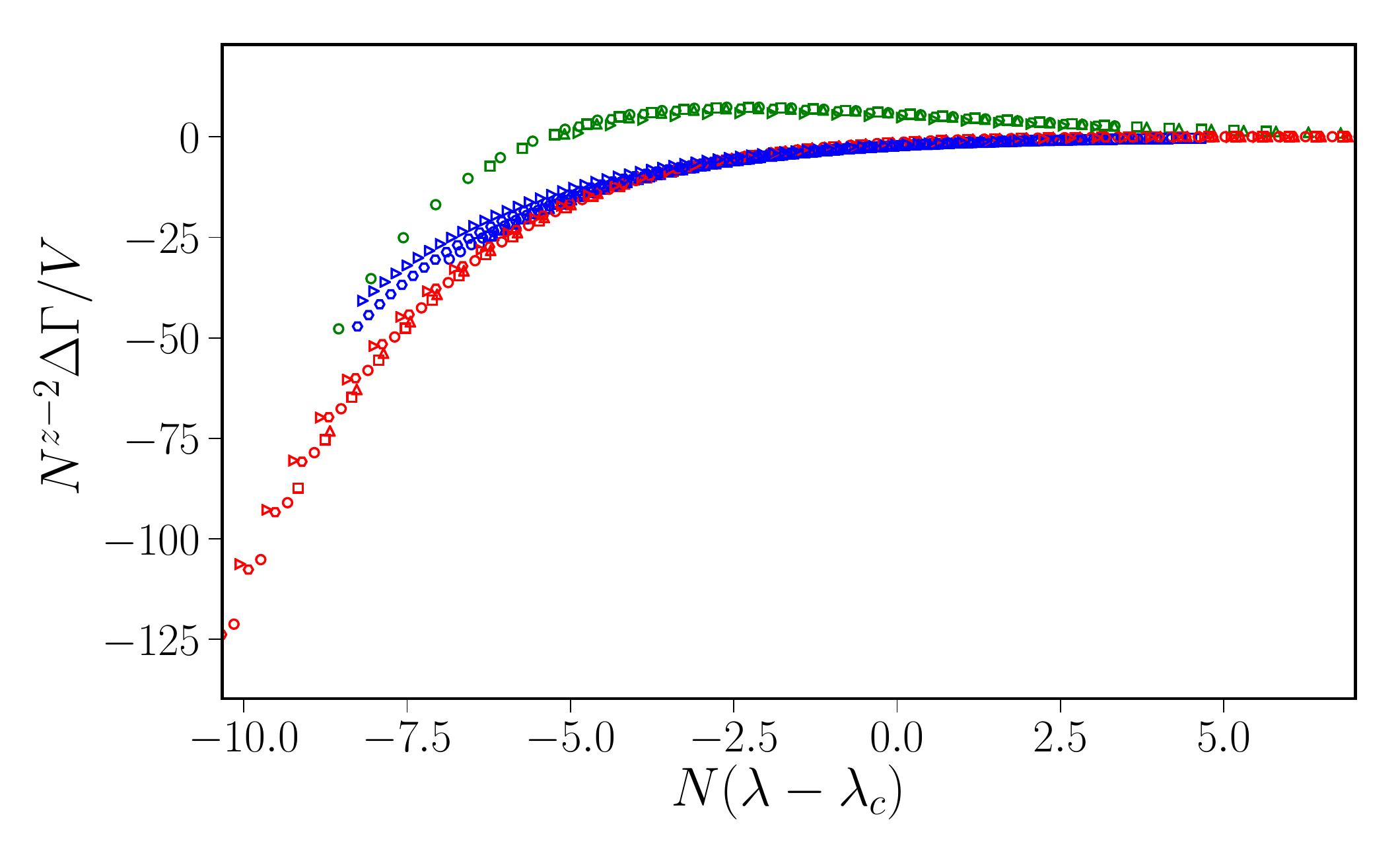}
        \put(0.1,60){$b)$}
    \end{overpic}\begin{overpic}[width=.33\textwidth]{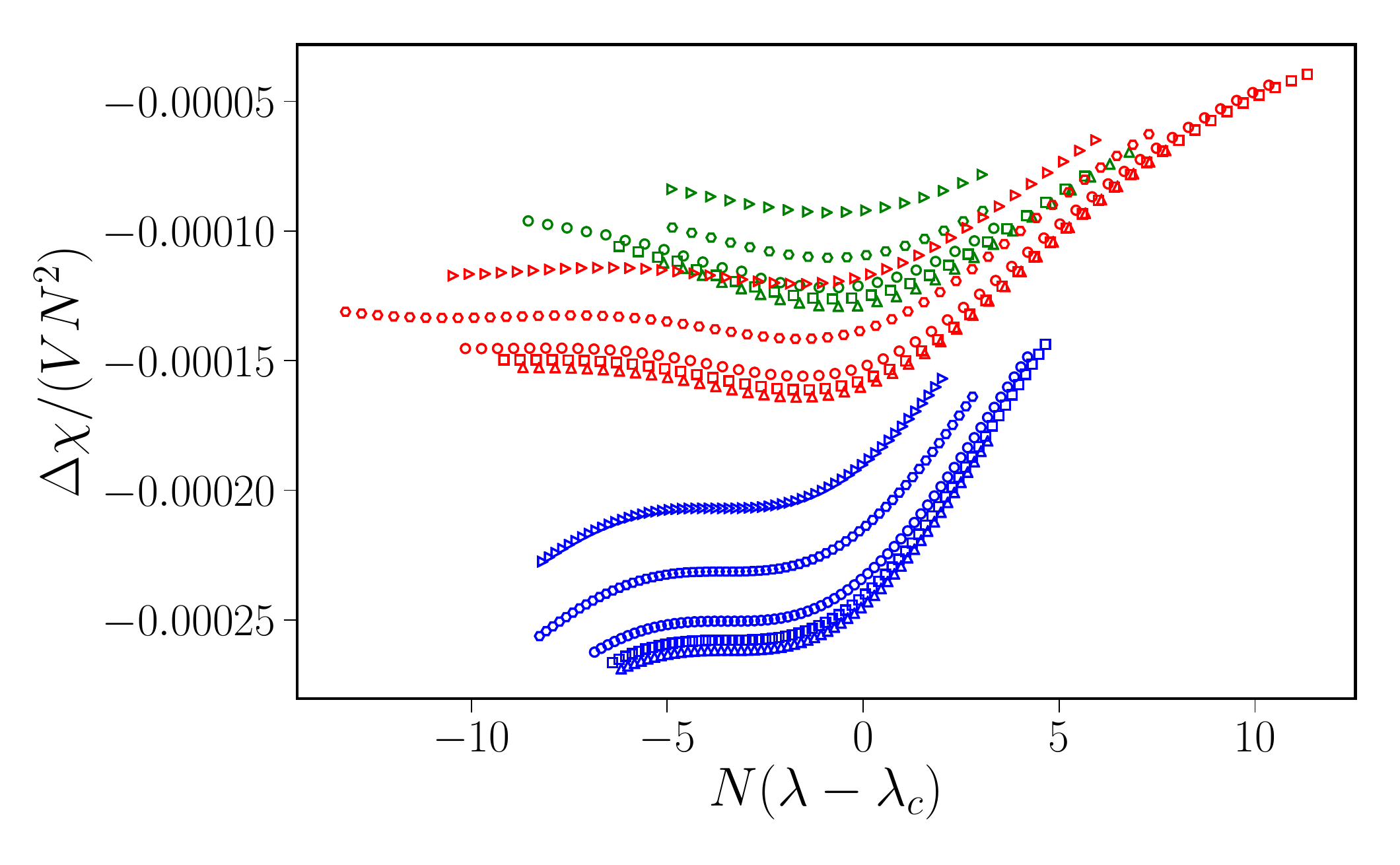}
        \put(0.1,60){$c)$}
    \end{overpic}
    \caption{(Color online) a) Even in the presence of interactions, $\Gamma$ still scales onto the same curve with $\lambda_c\approx 2.0 + 0.23 V$, $z=1.575$, and $\nu=1$. The legend applies to a), b), and c): each system size $N$ corresponds to a different color, and different symbols correspond to different $V$. In b), c), it is clear that the change in $\Gamma$ and $\chi_F$ is proportional to $V$ showing that it is likely irrelevant since such proportionality would break-down at larger $N$ if it were relevant. There is no universal function collapse in the b), c) because there are at least two irrelevant directions that $V$ contributes to. Because the collapse is worse for $\Delta \chi$ than for $\Delta \Gamma$, it appears that $\chi$ suffers from more finite-size effects. }
    \label{fig:dmrghf}
\end{figure*}

The PBC allow us to reliably reach a maximum system size of $\lesssim 200$. Since the best rational approximation's denominator, $N_k$, grows exponentially, it is difficult to find $\beta$ which provide enough accessible system sizes. The most dense denominators occur for the golden ratio, $\beta_{11}$, but due to its three universality classes at half-filling, there are only two system sizes for each of the three classes with $10<N_k<200$. We instead focus on the following three cases: $\beta=1/\sqrt{2}=\beta_{21}$ at half-filling, $\beta=``0$'' at half-filling, and $\beta=\beta_{11}$ at the commensurate filling $2\beta_{11}-1$.

For  half-filling and  $\beta = 1/\sqrt{2}$, all values of $N$ belong to the same universality class and we can easily access three system sizes. In Fig.~\ref{fig:dmrghf}, we notice that $\chi$, $\Gamma$ still collapse onto the same curves with the same exponents. We determine $\lambda_c\approx2.0+\Delta \lambda$ where $\Delta \lambda$ is how much the peak of $\chi_F$ shifts for the largest $N$ shown. 
Since there is no change in the exponent and the curves remain essentially the same, we suspect that $V$ is irrelevant or marginal. We can attempt to estimate the exponent of the irrelevant direction via a finite-size scaling analysis.

We assume the scaling hypothesis of a quantity $X$ to write
\begin{equation}
\begin{aligned}
    X &= f(|t|N^{1/\nu}, u_1N^{y_1},u_2N^{y_2}, ...) \\&= f(|t|N^{1/\nu}, 0, 0, ...) \\&+ \sum_{i} u_iN^{y_i}f_i(|t|N^{1/\nu}, 0, 0, ...)  
\end{aligned}
\end{equation}
where $f(x_0,x_1,x_2,...)$ and $f_i = \partial f/\partial x_i$. Since each $u_i$ is a linear combination of $\lambda, V$ and potentially other parameters (if the RG procedure is not closed), then we will not be able to easily collapse the functions onto a universal curve if $u_i\ne 0$ for $i>1$ or if $|y_2|\not \gg |y_1|$. 

The typical procedure to estimate irrelevant exponents would have us fit $\chi_{F,\text{max}} \sim N^{2/\nu}(1+a_1 N^{y_1})$. Due to the small number of accessible system sizes, we instead attempt to see if the curves completely collapse. When we attempt such a collapse in Fig.~\ref{fig:dmrghf}, a universal curve does not seem to emerge. However, for small $V$, we are able to obtain a decent collapse of $\Delta \Gamma = \Gamma(V)-\Gamma(0)$ at fixed $N$ onto the same curve as a function of $V$. This tells us that our results are not suffering from numerical issues as otherwise they would not be proportional to $V$. We note that the slight non-linearity in $V$ of $\Delta \chi_F$ is caused by finite-size corrections.  Since the data does not collapse well onto a single universal curve, we conclude that the interaction term contributes to at least two irrelevant directions. We would therefore need at least four points to fit the $\chi_{F,\text{max}}$ data to estimate the most relevant irrelevant direction, but such an analysis would not be very conclusive.

\begin{figure*}
    \centering
    \begin{overpic}[width=.33\textwidth]{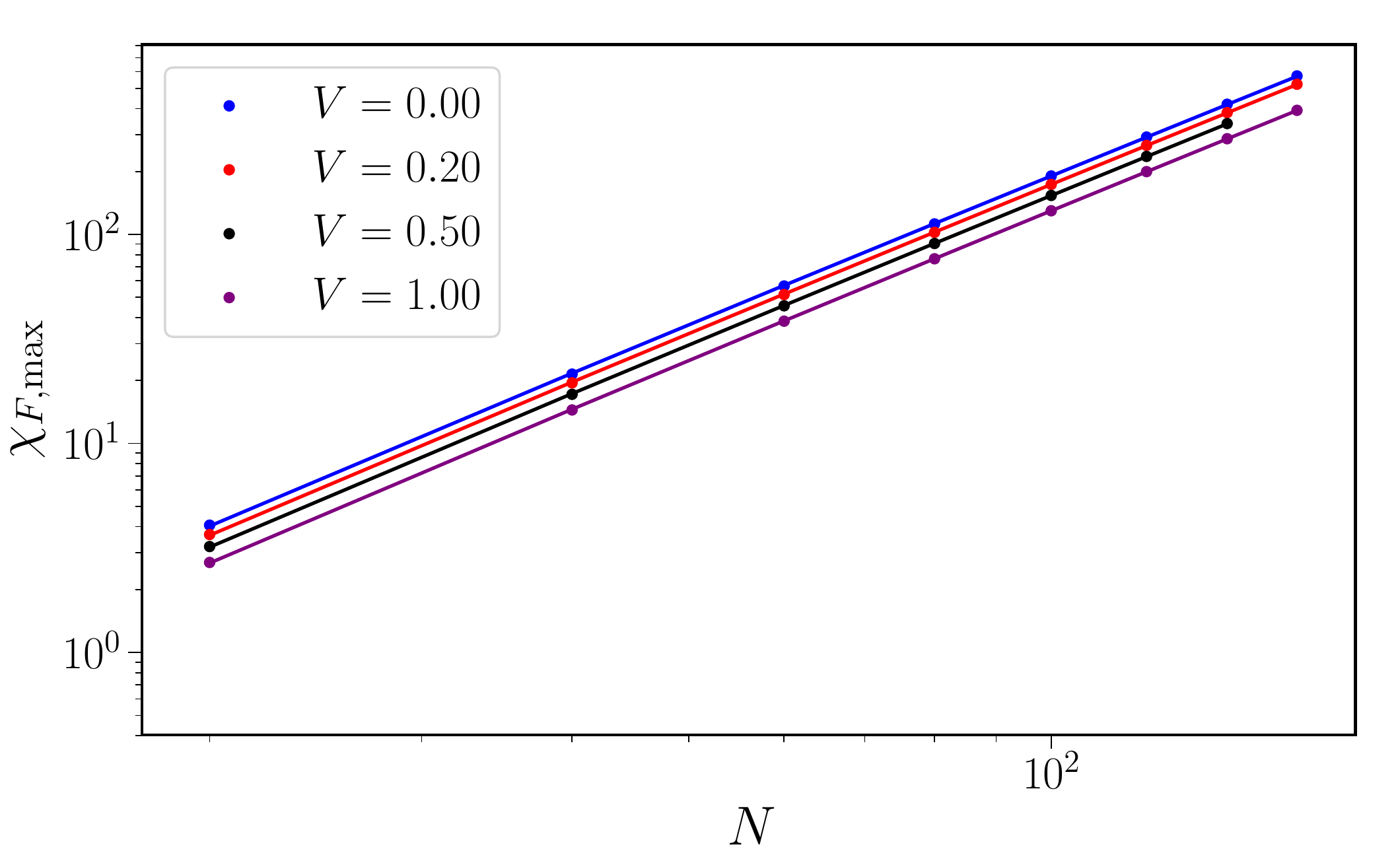}
        \put(0.1,60){$a)$}
    \end{overpic}
    \begin{overpic}[width=.33\textwidth]{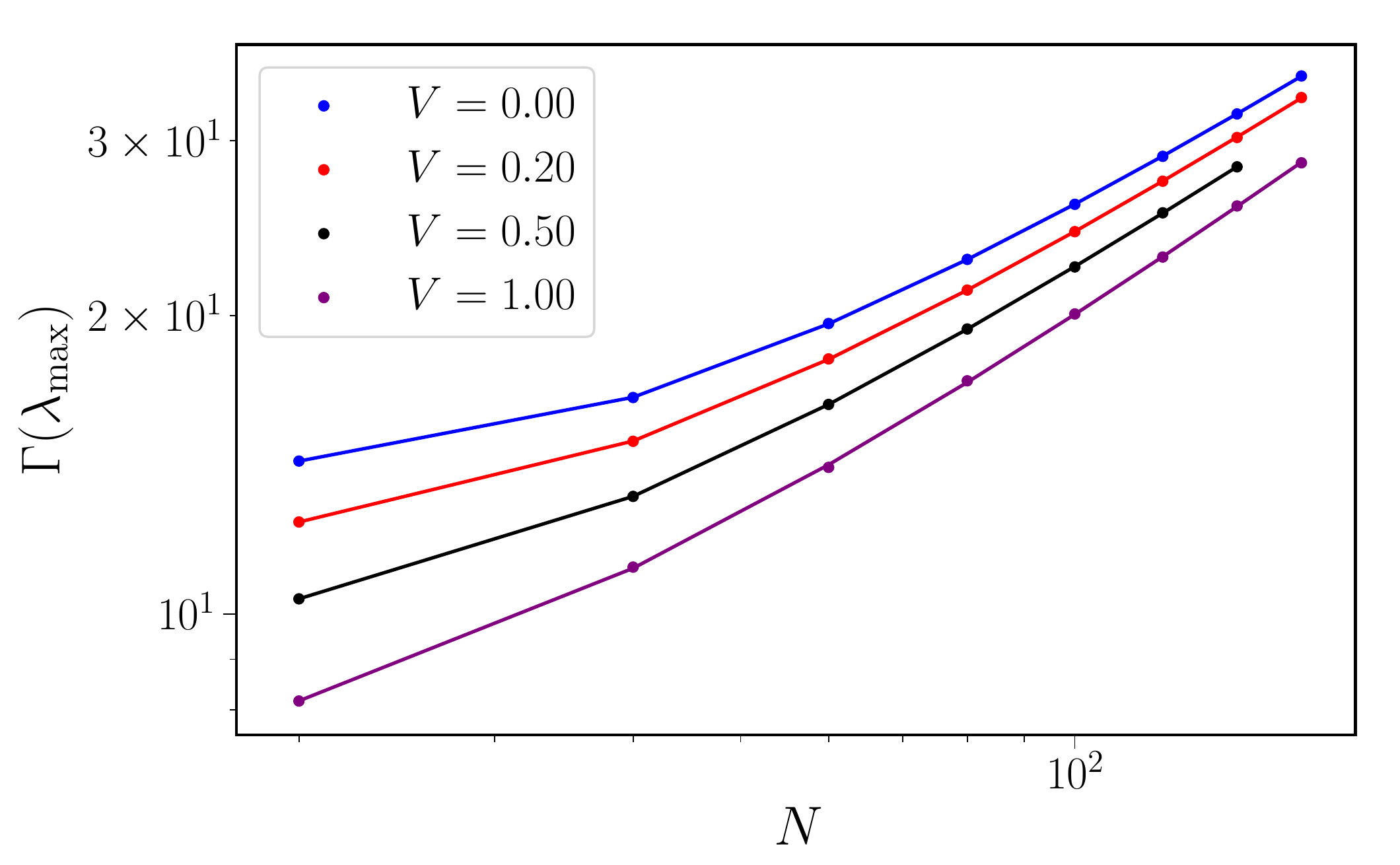}
        \put(0.1,60){$b)$}
    \end{overpic}\begin{overpic}[width=.33\textwidth]{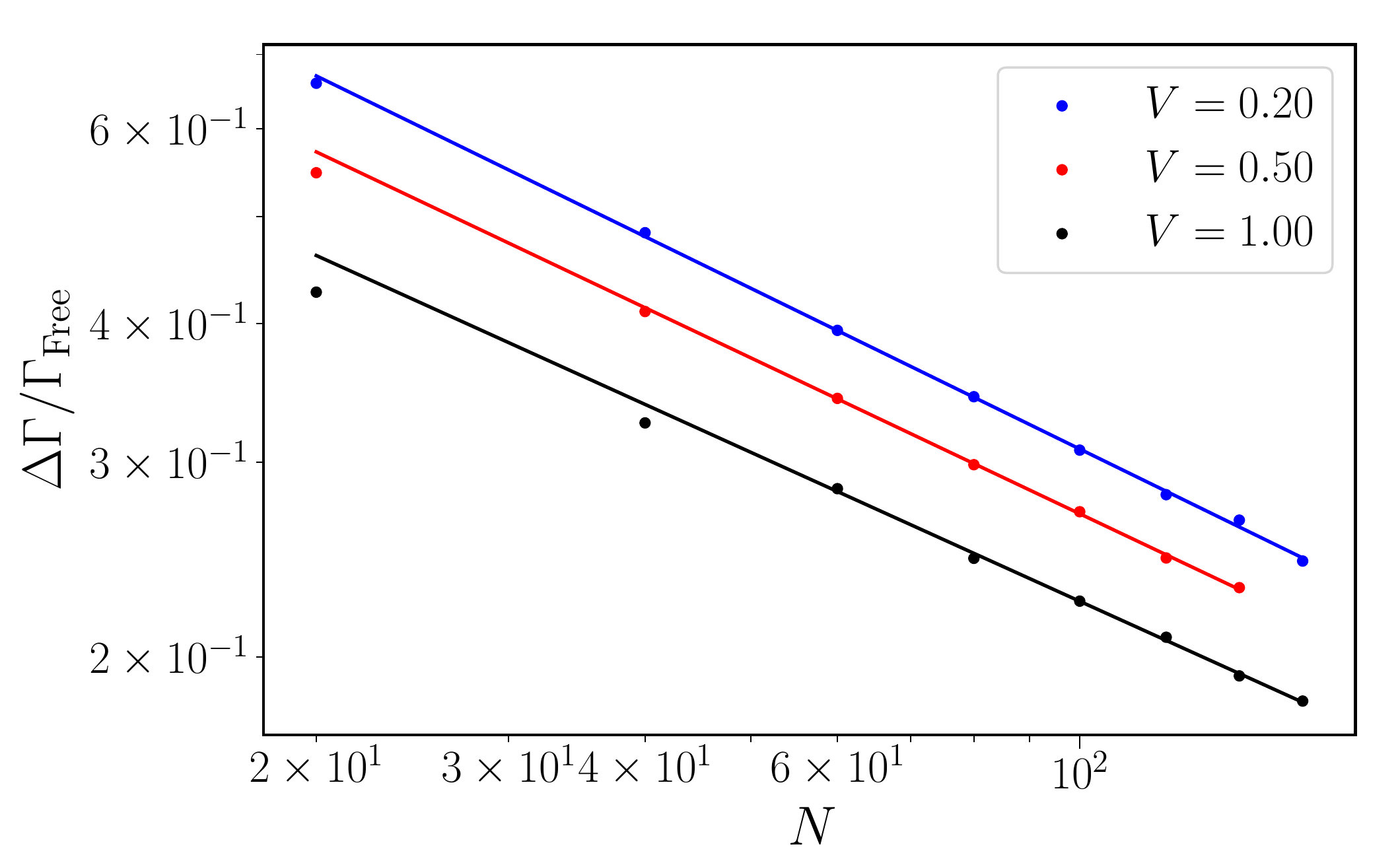}
        \put(0.1,60){$c)$}
    \end{overpic}
    \caption{In a) we plot the maximum of the fidelity susceptibility ocurring at $\lambda=\lambda_\text{max}$. The lines indicate a fit to $\chi_{F,\text{max}}\sim N^\mu \log(N)$ and $\mu\approx 2.1$ for all interactions. In b) we plot $\Gamma(\lambda_\text{max})$ and fit $\Gamma(\lambda_\text{max})\sim N^{2-z}(1+a_1 N^{y_1})$, which all yield $z\approx 1.25$ as in the free case. Part c) depicts $(\Gamma(V,\lambda_{\text{max}})-\Gamma(0,\lambda_\text{max})/\Gamma(0,\lambda_{\text{max}}) \equiv \Delta \Gamma/\Gamma_\text{free} \sim N^{-y_V}$ giving us $y_V \approx 0.44-0.48$ for all $V$ when we exclude points with $N<60$. }
    \label{fig:beta0Vne0}
\end{figure*}
With the standard $\beta \not \in \QQ$ analysis, there exists no good choice of $\beta$ that allows for enough accessible system sizes with this current analysis. 
To access more system sizes, we can consider $\beta = \text{``0''}$ by taking $\beta = 1/N$ for any $N$, as first discussed in Ref.~\onlinecite{Thakurathi2012}. This parameter choice leads to a very similar transition at half-filling in that, for $V=0$, $\nu=1$, $\lambda_c =2$, and $z=1.245$ (see Fig.~\ref{fig:betato0transition} for the free case).\footnote{Ref.~\onlinecite{Thakurathi2012} estimates that the width of the fidelity susceptibility scales with $N^{0.7}$ and the peak scales with $N^{2.25}$. The peak seems to scale with an exponent $\mu>2$, but we find that with a logarithmic correction taken into account, $\mu \approx 2.1$, instead of $\mu=2.25$. Additionally, the collapse of $\Gamma$ strongly suggests $\nu=1.0$, and, as $N$ gets larger, the fidelity susceptibility curves seem to also collapse better and better onto a curve with width scaling with $N$.} To keep the curves within the same universality class at half-filling, we choose $N$ that are divisible by four with $\phi=0$. Nevertheless, many more system sizes are accessible.\footnote{If we choose $\phi=\pi/2$, the universality classes for all even $N$ are the same, as is predicted by the Diophantine equation.}

To do the finite-size analysis, we compute $\chi(\lambda_\text{max})$ and $\Gamma(\lambda_\text{max})$ where $\lambda_\text{max}$ is the peak of the fidelity susceptibility (determined with a cubic interpolation of the points at which we performed DMRG). We first observe in Fig.~\ref{fig:beta0Vne0} that $\chi_{F,\text{max}}\sim N^\mu\log(N)$ with $\mu\approx 2.1$ and $\Gamma(\lambda_\text{max})\sim N^{2-z}(1+aN^{y_1})$ with $z\approx 1.2-1.3, y_1\approx1.1-1.3$ for all cases including the free case. Due to the nicer collapse of $\Gamma$ in the $\beta_{21}$ case and to isolate the effect of $V$, we consider the quantity $(\Gamma(V,\lambda_{\text{max}})-\Gamma(0,\lambda_\text{max})/\Gamma(0,\lambda_{\text{max}}) \equiv \Delta \Gamma/\Gamma_\text{free} \sim N^{-y_V}$ and perform a scaling analysis. Using the scaling hypothesis, we expect this to be the most relevant irrelevant direction that $V$ contributes to. Our analysis gives $0.44 < y_V < 0.48$ when we use system sizes with $N>60$ for the fit. When we perform a similar analysis on $\Delta \chi_F/\chi_{F,\text{free}}$, we get $y_V \sim 0$ for low $V$, but we know from Fig.~\ref{fig:dmrghf}c that finite size effects influence this quantity more. Regardless, the above very much suggests that $V$ is irrelevant or marginal.

Finally, we turn to the commensurate filling of $2\beta_{11}-1$ for $\beta=\beta_{11}$.
Because of the irrelevance of a given interaction $V$, we expect that at incommensurate filling, the system flows towards the $\lambda=2, V=0$ critical point. However, it is unclear if that is true when the transition is at $\lambda=0$.
The authors of Ref.~\onlinecite{Schuster2002} studied a localization-delocalization transition at commensurate filling when $V>\sqrt{2}$ coming from the Peierl's type resonance we discuss above. They were unable to get a scaling collapse in $\lambda$, which we focus on.

\begin{figure*}
    \centering
    \begin{overpic}[width=.45\textwidth]{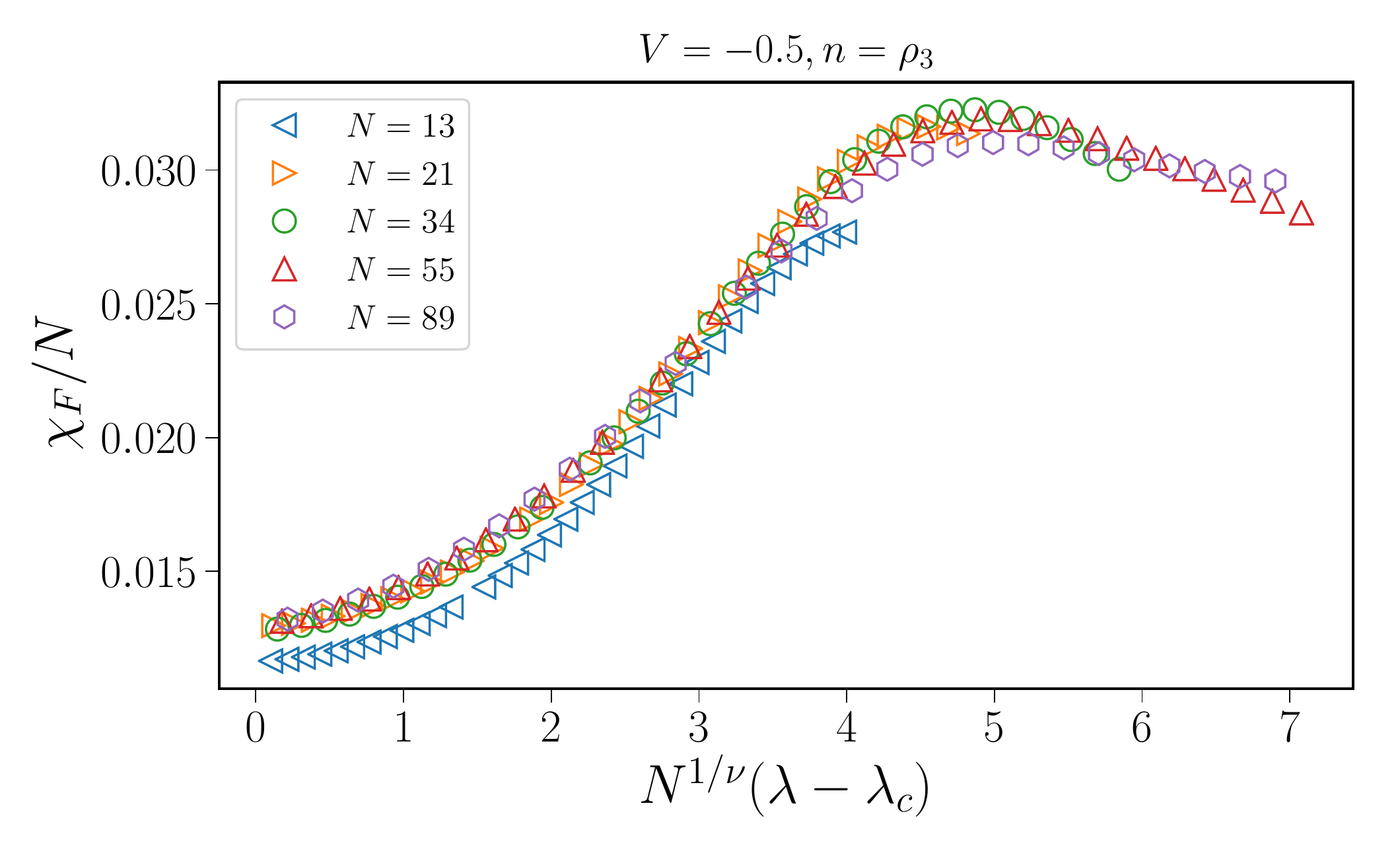}
        \put(0.1,60){$a)$}
    \end{overpic}
    \begin{overpic}[width=.45\textwidth]{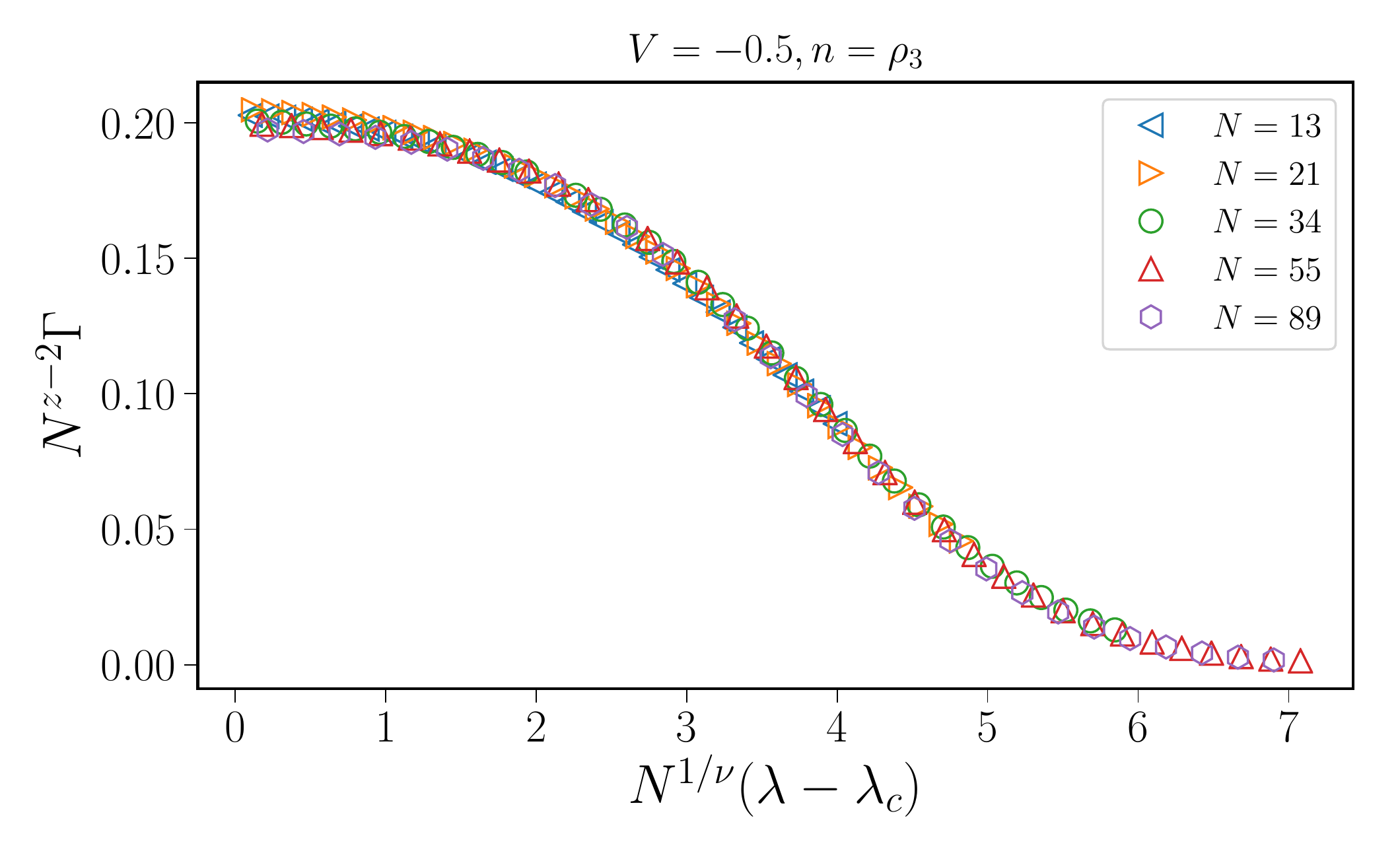}
        \put(0.1,60){$b)$}
    \end{overpic}
    
    \begin{overpic}[width=.45\textwidth]{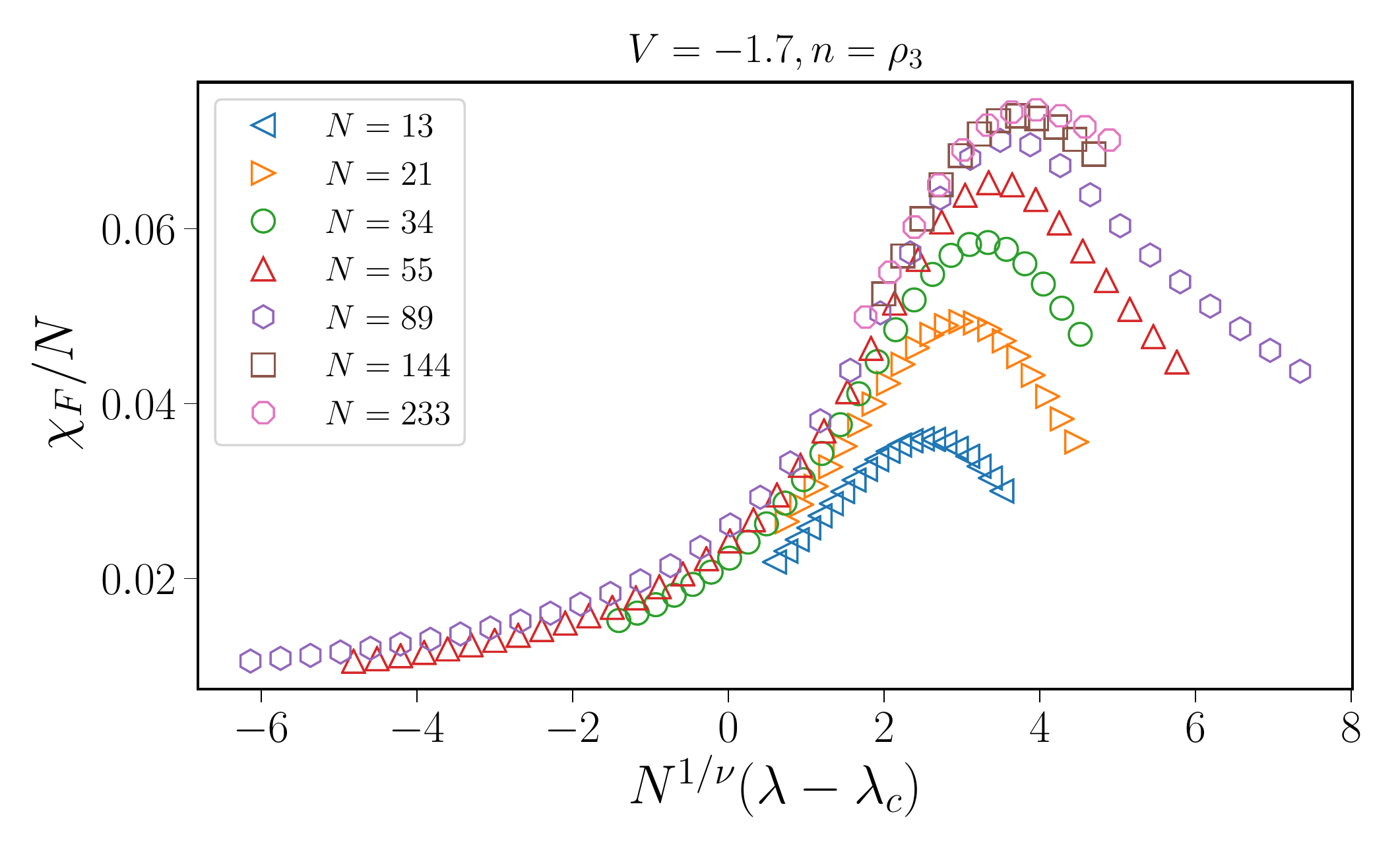}
        \put(0.1,60){$c)$}
    \end{overpic}
    \begin{overpic}[width=.45\textwidth]{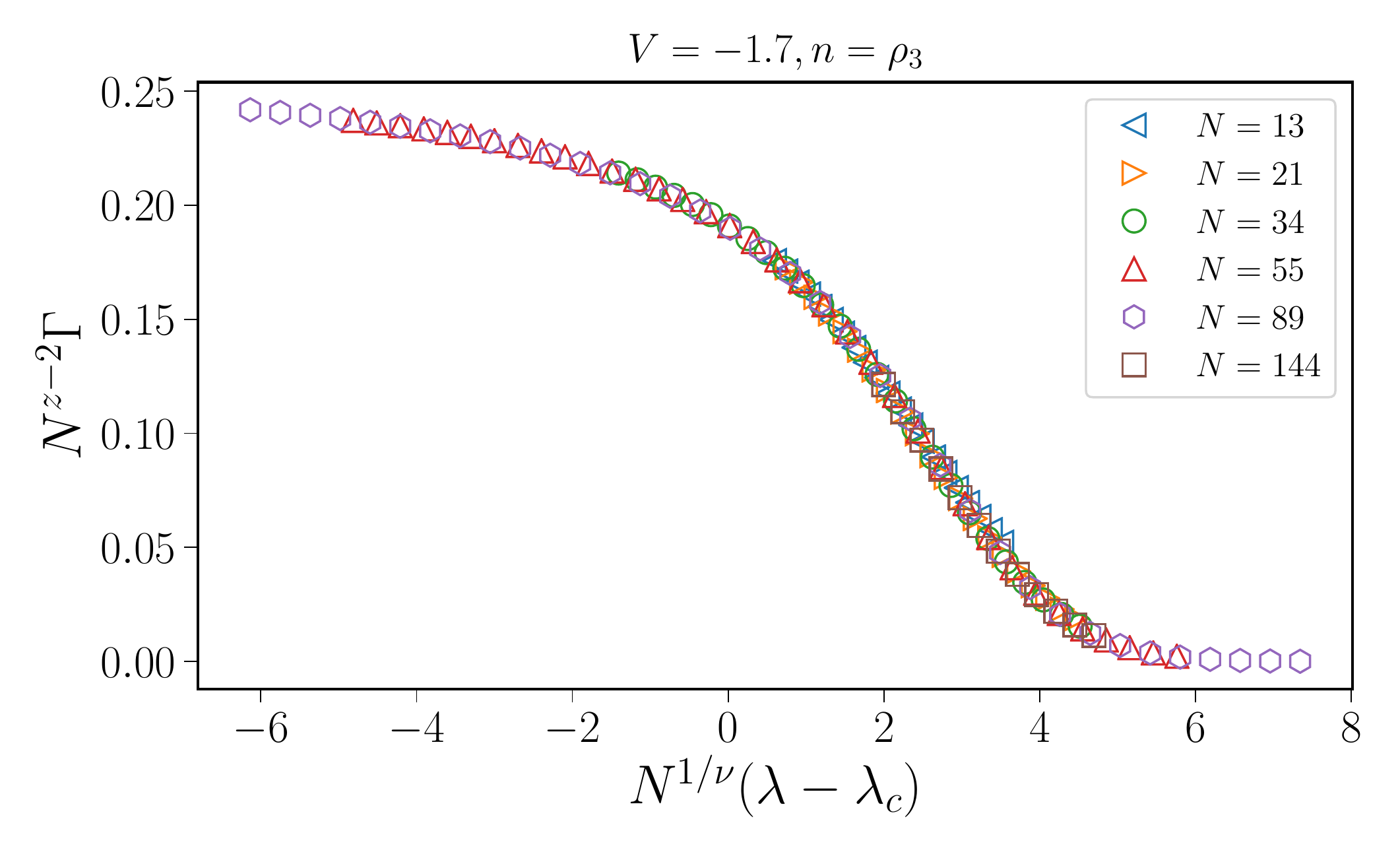}
        \put(0.1,60){$d)$}
    \end{overpic}
    \caption{We plot the scaling collapse of $\Gamma$ [in b) and d)] and $\chi_F$ [in a) and c)] for commensurate filling $\rho_3 = 2\beta_{11}-1$ and two different interaction strengths, where the exponents are determined using $\Gamma$ only. In the free case, $\nu=2,z=1$ and $\lambda_c=0$, and the scaling collapse for $\Gamma$ in b) and d) gives $\nu\approx 2$ and $z\approx 1$ with $\lambda_c \approx 1$ for $V=-1.7$ and $\nu\approx 2.5$, $z\approx 1$, and $\lambda_c \approx 0.1$, so the exponents have not changed much. Although numerically determined, $\Gamma$ close to the transition should be accurate and less prone to finite-size effects than $\chi_F$. However, finite-size effects are the largest source of error. }
    \label{fig:CommFillDMRGGR}
\end{figure*}

If it is similar to the incommensurate case, we expect that the transition will shift away from the free point, but the exponent will stay the same. Since $\nu=2$, the fidelity is less useful as a gauge for the location of the transition as the fidelity does not grow super-extensively. We can, however, attempt a finite-size scaling allowing $\lambda_c,z,$ and $\nu$ to vary and minimize the following quantity \cite{newman1999}
\begin{equation}
\begin{aligned}
    \sigma^2&=\frac{1}{2\Delta x}\int_{x_0-\Delta x}^{x_0+\Delta x} dx \langle g(x)^2\rangle - \langle \tilde g(x)\rangle^2\\
    g(N^{1/\nu}(\lambda-\lambda_c)) &= N^{z-2}\Gamma(\lambda),
\end{aligned}
\end{equation}
where we use cubic interpolation of the values of $g(x)$ with no explicit evaluation and where $\langle \cdot \rangle$ is an average over the available $N$. The results of this fitting are shown in Table~\ref{tab:intCritProps} and, for $V=-0.5$ and $V=-1.7$, Fig.~\ref{fig:CommFillDMRGGR}.

 In contrast to Ref.~\onlinecite{Schuster2002}, the exponents seem to be roughly the same as the ones expected in the free case, namely $\nu=2, z=1$ when we perform the finite-size fitting. It should be noted that there is no clear way to estimate errors on our values because the dominant error would come finite-size effects which we are neglecting. Adding in these effects would allow for too many parameters to meaningfully constrain the exponents.

Although the transition is still controlled by the same RG fixed point, there is now a finite range of $\lambda \in (-\lambda_c,\lambda_c)$ where the wave function is extended. Additionally, there are rather large finite size effects in the fidelity susecptibility at $V=-1.7$, which seem to decrease for $V=-0.5$ (see Fig.~\ref{fig:CommFillDMRGGR}). This suggests that $V$ is irrelevant or marginal in this case. It is not feasible to perform the same analysis for $\beta = \text{``0''}$ because the peak of the fidelity susceptibility is not as reliably close to the transition due to it growing only extensively. 

\begin{table}[]
    \centering
    \begin{tabular}{c|c||c}
        $\beta,\rho$ & $V$ & $\lambda_c$ \\
        \hline
        \hline
        $\beta_{21},1/2$ & 0.05 & $2.011$ \\
         & 0.1 & $2.023$\\
         & 0.2 & $2.046$\\
         & 0.5 & $2.119$ \\
         & 1.0 & $2.249$ \\
         \hline
         \hline
        ``$0$''$,1/2$ & 0.2 & 2.2 \\
        & 0.5 & 2.5\\
        & 1.0 & 3.0 \\
        \hline \hline
        $\beta_{11},2\beta_{11}-1$ & -0.5 & (0.05-0.15,0.4,-1) \\
        & -1.3 &  (0.5, 0.4, -1)\\
        & -1.5 &  (0.8, 0.4-0.5, -1)\\
        & -1.7 &  (1.05, 0.51, -0.95)\\
    \end{tabular}
    \caption{For the first two $\beta$'s, $\lambda_c$ is determined from the shift in $\chi_F$ for the largest system size probed. 
    In the last case, it comes from finite-size fitting for all $(\lambda_c, 1/\nu, z)$, where all three numbers or a range are reported. Significant figures are chosen to capture the range of values the minimization converges to. The dominant source of error is finite-size effects, which, as mentioned in the text, are difficult to account for or accurately estimate. Since the exponents do not change significantly in any case we considered, we conclude that the Diophantine equation determines the universality class even in the presence of interactions.}
    \label{tab:intCritProps}
\end{table}

Finally, we note that how the transition depends on $V$ is highly dependent on $\beta$. We saw that, at half-filling, $\Delta \lambda \approx 0.23 V$ in the case of $\beta_{21}$ and that $\Delta \lambda \approx V$ for $\beta = \text{``0''}$ whereas $\Delta \lambda \approx 0$ for $\beta = \beta_{11}$ \cite{Schuster2002}.

\section{Conclusions} \label{sec:conclusion}

By analyzing the Diophantine equation that naturally arises in the Aubry-Andr\'e model, we have found that it accurately determines the dynamic-critical exponent $z(\beta,\rho)$ for the incommensurate ratio $\beta$ and filling factor $\rho$. This analysis yielded non-trivial relationships between different $\beta$ and different $\rho$ that shows the universality depends on more than just the continued fraction expansion of $\beta$ as is seen in the single-particle case.\cite{Szabo2018,Hashimoto1992}. The dynamic-critical exponent is related to the multifractal properties of the system as it describes how different sections of the energy-spectrum scale with system size. The major results testing the Diophantine connection between critical exponents in the non-interacting case is summarized in Table~\ref{tab:exponents} and explicit examples can be seen in Figs.~\ref{fig:SR} and \ref{fig:Commfillfree}. As noted by Ref.~\onlinecite{Tang1986}, the low temperature specific heat should go like $T^{1/z}$ making the universality, in principle, measurable. 

Such a relationship may seem contrived or pathological, but we have provided evidence that the exponents are nearly insensitive to the simplest form of interactions (see Table~\ref{tab:intCritProps} for $\beta_{11}$). The degree of irrelevance of the interaction is measured for $\beta=\text{``0''}$, but the results are inconclusive. We expect that for large enough interactions, the transition will become first-order or cease to exist as large enough $V$ will induce a charge-density wave state \cite{Naldesi2016}. 

\section{Acknowledgements}

The authors were supported by NSF DMR-1918065  and an NSF graduate fellowship (T.C. and J.E.M.), TIMES at Lawrence Berkeley National Laboratory supported by the U.S. Department of Energy, Office of Basic Energy Sciences, Division of Materials Sciences and Engineering, under Contract No.\ DE-AC02-76SF00515 and by DFG research fellowship MO 3278/1-1 (J.M.), and a Simons Investigatorship (J.E.M.).

\bibliography{IntAA.bib}

\section*{Appendix A: Operator algebra derivation of scaling quantities.}

We will derive the numeric expressions we are using. We consider the equation
\begin{equation} 
    \chi_{F,2+2r} = \sum_{n\ne 0} \frac{ | \langle \Psi_n | H_I | \Psi_0 \rangle|^2 }{(E_n - E_0)^{2+2r}} \sim N^{\mu +2zr}
\end{equation}
for the generalized fidelity susceptibility. Since we are considering $\lambda$ as the tuning parameter, $H_I =  \sum_i h_i n_i$. We can switch bases and rewrite
\begin{equation}
    H = \sum_{i,j}c_i^\dagger H_{ij} c_j = \sum_{i}\gamma_{i}^\dagger \gamma_i \lambda_i
\end{equation}
for $\gamma_i = S_{ij}c_j $ since we diagonalize $H_{ij} = S_{ik}^\dagger \Lambda_{kl}S_{lm}$. The ground state with $N_F$ particles will be 
\begin{equation}
    |\Psi_0 \rangle  = \gamma_{N_F}^\dagger \gamma_{N_F-1}^\dagger\cdots \gamma_1^\dagger |0 \rangle
\end{equation}
where the energies $\lambda_i$ are sorted from least to greatest. In this basis, we can write:
\begin{equation}
    H_I = \sum_{i,j,k} h_i S_{ji}^\dagger S_{ik} \gamma_j^\dagger \gamma_k
\end{equation}
which clearly only drives transitions between the ground state and states where we have excited one of the particles to a higher state. Therefore:
\begin{equation}
    \chi_{F,2+2r} = \sum_{j>N_F,k\le N_F} \frac{ |\sum_i S_{ij}^* S_{ik}h_i|^2}{(\lambda_j - \lambda_k)^{2+2r}}
\end{equation}
There are $N_F(N-N_F)/2$ states that contribute to this sum, so the operator scales as $\mathcal O(N^3)$ if $N_F\sim N$. Because diagonalizing the matrix is $\mathcal O(N^3)$ anyway, it doesn't hurt the overall scaling. 

To compute $\Gamma = N^2\partial E/\partial \theta$, we need to do perturbation theory where the perturbation is $H_I = P_F(e^{i\theta}-1)c_N^\dagger c_1 + P_F(e^{-i\theta}-1)c_1^\dagger c_N$. We need to compute the coefficient of $\theta^2$. In addition to a term like in $\chi_{F,2+2r}$, there is an additional term from first order perturbation theory where we have Taylor expanded $e^{i\theta}-1$ and kept to second order. Therefore
\begin{equation}
\begin{aligned}
\Gamma &= N^2 \left[ \sum_{i} \frac{P_F}{2}\left(S_{iN}^\dagger S_{1i} + S_{i1}^\dagger S_{Ni}\right) \right. \\
 &\left.+ \sum_{j>N_F, k\le N_F} \frac{| S_{jN}^\dagger S_{1k}- S_{j1}^\dagger S_{Nk}|^2}{\lambda_j-\lambda_k}\right]
 \end{aligned}
\end{equation}
which is $\mathcal O(N^2)$.

\section*{Appendix B: Diophantine equation manipulations}\label{app:Diophantine}

We will derive that $(\beta_{nn},\rho)$ and $(\beta_{nn},\rho/\beta_{nn})$ belong to the same universality class rigorously under our conjecture (i.e. they have the same value of $Q_k/N_k$ for $k\gg 1$). We can use similar manipulations to make the hand-wavy analyses in the main text [such as Eq.~\eqref{eq:beta21beta22univ} and Eq.~\eqref{eq:beta_nn_fill}] more rigorous.

Suppose that we have a solution $Q_k,P_k$ to the Diophantine equation $Q_k M_k - P_k N_k = N_F $. Then, we can use that $M_k/N_k = N_{k-1}/(nN_{k-1}+M_{k-1})$ for $\beta_{nn}$ to write
\begin{equation}
\begin{aligned}
    N_F \frac{(nN_{k-1}+M_{k-1})}{N_k}&=Q_k N_{k-1} -P_k (nN_{k-1}+M_{k-1}) \\
    &=(Q_k-nP_k)N_{k-1}-P_kM_{k-1}.
\end{aligned}
\end{equation}
Notice that, in the limit that $k\to \infty$, $N_F=N_k\rho$ and $N_F' = N_F(nN_{k-1}+M_{k-1})/N_k = N_{k-1}\rho/\beta$ where we used $\beta_{nn}= M_{k}/N_{k}$. Therefore, the solutions to the Diophantine equation at these two fillings are related. It now suffices to show that $P_k/N_{k-1}=Q_k/N_k$ as $k\to \infty$. Recall that $N_{k-1}/N_k \to \beta_{nn}$, and, if $P_k$ grows extensively, then the Diophantine equation reveals that $M_k/N_k-P_k/Q_k=N_F/(N_kP_k)\to 0$, so $P_k/Q_k \to \beta_{nn}$ as well.

\section*{Appendix C: Details of the DMRG calculations}

As discussed in the main text,  DMRG computations for (A)PBC require larger bond dimension than for open systems. In order to ensure convergence, we use the following procedure. 
Let $H(\lambda)$ be the Hamiltonian at the parameter value $\lambda$ and $\Psi(\lambda)$ is the ground state wave function achieved by performing DMRG. For the fidelity, we perform $n_1$ sweeps on $\Psi(\lambda)$ with a maximum bond dimension of $M_1$. We then start doing two sweeps with a maximum bond dimension at $M_1 + M_2 \lfloor n_\text{sweeps}/2 \rfloor$ for $n_\text{sweeps}$ the total number of sweeps on $\Psi(\lambda)$ and $\Psi(\lambda\pm \delta \lambda)$ (using $\Psi(\lambda)$ as the initial guess). After the two sweeps, we compute 
\begin{equation}
    \chi_F =  -2\frac{\ln\left\{\frac{\left[\langle \Psi(\lambda) | \Psi(\lambda+\delta \lambda) \rangle + \langle \Psi(\lambda) | \Psi(\lambda-\delta \lambda) \rangle \right]}{2}\right\}}{\delta \lambda^2} + \mathcal O(\delta \lambda^2)
\end{equation}
and compare with the previously computed value. Once the relative change is below $\epsilon$, we consider it converged.

To compute $\Gamma$, we follow a similar procedure but we are doing sweeps on $\Psi(\lambda), \Psi(\lambda,\theta=\theta_0), \Psi(\lambda,\theta=2\theta_0),\Psi(\lambda,\theta=3\theta_0)$, and we compute $\Gamma$ as
\begin{equation}
\begin{aligned}
    \Gamma &=N^2 \frac{-245 E(0) +270 E(\theta_0) -27 E(2\theta_0)+2E(3\theta_0)}{ 90 \theta_0^2} \\& + \mathcal O(\theta_0^6),
\end{aligned}
\end{equation}
where $E(\theta)$ is the energy of $\Psi(\lambda,\theta)$.

We generally use parameters $(n_1,M_1,M_2,\epsilon,\delta\lambda,\theta_0)=(6,300,100,10^{-5},0.001,\pi/30)$. We compare the values of $\chi_F$ and $\Gamma$ computed with the above formula as well as those with lower-order finite-difference expressions to ensure reasonable accuracy. We also calculated them in the $V=0$ case and found good agreement.

\section*{Appendix D: Universality class at 1/6 filling.}

The results presented in this section are solely based on the Diophantine equation. As mentioned in the main text, the two universality classes are considered the same if the same sequence of values of $q_k=|Q_k|/N_k$ appears, and this sequence has a period of $p$. We omit the sequence of $q_k$ for clarity.

\begin{table}[]
    \centering
    \begin{tabular}{c|c}
          Universality class & $p$ \\
         \hline
         \hline
        $\beta_{11},\beta_{12},\beta_{13},\beta_{14},\beta_{15}$ & 12 \\
        \hline
                \hline
        $\beta_{21},\beta_{23},\beta_{25},\beta_{27}$ & 8\\
                \hline
        $\beta_{22},\beta_{26},\beta_{28}$ & 4\\
                \hline
        $\beta_{24}$ & 8\\
        \hline
        \hline
        $\beta_{31},\beta_{34},\beta_{37}$ & 3\\
                \hline
        $\beta_{32},\beta_{35}$ & 3\\
                \hline
        $\beta_{33},\beta_{36}$ & 6\\
        \hline
        \hline
        $\beta_{41},\beta_{45}$ & 8\\
                \hline
        $\beta_{42},\beta_{48}$ & 8\\
                \hline
        $\beta_{43},\beta_{47}$ & 8\\
                \hline
        $\beta_{44},\beta_{46}$ & 4\\
    \end{tabular}
    \caption{The $\beta$'s within the same cell belong to the same universality class, based on the analysis of the Diophantine equation alone. The value of $p$ is given, but we omit the sequence of $|Q_k|/N_k$ which differentiates those classes with the same value of $p$. Note that when $\beta_{pm}$ has $p$ with a common divisor to $6=1/\rho$ the universality classes depend on more than just $p$, which determines the asymptotic continued fraction expansion of $\beta_{pm}$.}
    \label{tab:onesixthuniversality}
\end{table}

We find the results in Table~\ref{tab:onesixthuniversality}. Notably, the universality class for $\beta_{1m}$ for $m\in \{1,2,3,4,5\}$ are all the same, whereas there are different classese for $\beta_{2m},\beta_{3m}$, and $\beta_{4m}$ for a filling of $\rho=1/6$. This evidence supports the notion that $\beta_{pm}$ can split into different universality classes at a filling of $1/q$ if $p$ and $q$ share a prime divisor, but the details are not obvious. It is not, for instance, that $m \text{ mod } p$ or $m \text{ mod } q$ determines the universality, based on Table~\ref{tab:onesixthuniversality}.

\end{document}